\documentclass{article}

\newif\ifacm
\acmfalse

\newif\ifpublic
\publictrue

\newcommand{\citerepo}{\ifpublic \cite{repo}\else [anonymous]\fi}

\usepackage[utf8]{inputenc}
\usepackage[T1]{fontenc}

\ifacm

\acmConference[CCS'19]%
  {Conference on Computer and Communications Security}%
  {November 11--15, 2019}%
  {London, UK}

\else 

\usepackage{graphicx}
\usepackage{hyperref}
\usepackage{amsmath}
\usepackage[margin=1.2in]{geometry}
\usepackage{subcaption}
\usepackage{authblk}

\fi 

\begin{document}

\title{Factoring semi-primes with (quantum) SAT-solvers}

\ifacm

\author{Michele Mosca}
\affiliation{%
  \department{Institute for Quantum Computing}
  \department{Department of Combinatorics \& Optimization}
  \institution{University of Waterloo}
  \country{Canada}
}
\affiliation{%
  \institution{Perimeter Institute for Theoretical Physics}
  \city{Waterloo}
  \country{Canada}
}
\affiliation{%
  \institution{Canadian Institute for Advanced Research}
  \city{Toronto}
  \country{Canada}
}
\affiliation{%
  \institution{evolutionQ Inc.}
  \city{Waterloo}
  \country{Canada}
}
\email{michele.mosca@uwaterloo.ca}

\author{Sebastian R. Verschoor}
\affiliation{%
  \department{Institute for Quantum Computing}
  \department{David R. Cheriton School of Computer Science}
  \institution{University of Waterloo}
  \country{Canada}
}
\email{srverschoor@uwaterloo.ca}

\setcopyright{none}

\else 

\author[1]{Michele Mosca\thanks{\href{mailto:michele.mosca@uwaterloo.ca}{michele.mosca@uwaterloo.ca}}}
\author[2]{Sebastian R. Verschoor\thanks{\href{mailto:srverschoor@uwaterloo.ca}{srverschoor@uwaterloo.ca}}}
\affil[1,2]{Institute for Quantum Computing, University of Waterloo, Canada}
\affil[1]{Department of Combinatorics \& Optimization, University of Waterloo, Canada}
\affil[2]{David R. Cheriton School of Computer Science, University of Waterloo, Canada}
\affil[1]{Perimeter Institute for Theoretical Physics, Waterloo, Canada}
\affil[1]{Canadian Institute for Advanced Research, Toronto, Canada}
\affil[1]{evolutionQ Inc., Waterloo, Canada}

\fi 

\date{\today}

\ifacm
\relax
\else
\maketitle
\fi

\begin{abstract}
The assumed computationally difficulty of factoring large integers forms the
basis of security for RSA public-key cryptography, which specifically relies on
products of two large primes or \emph{semi-primes}.
The best-known factoring algorithms for classical computers run in sub-exponential time.
Since integer factorization is in NP, one can reduce this problem to any NP-hard problem,
such as Boolean Satisfiability (SAT).
While reducing factoring to SAT has proved to be useful for studying SAT solvers,
attempting to factor large integers via such a reduction has not been found to be successful.

Shor's quantum factoring algorithm factors any integer in
polynomial time. Large-scale fault-tolerant quantum computers capable
of implementing Shor's algorithm are not yet available,
so relevant benchmarking experiments for factoring via Shor's algorithm
are not yet possible.
In recent years, however, several authors have attempted factorizations
with the help of quantum processors via reductions to NP-hard problems.
While this approach may shed some light on some algorithmic approaches
for quantum solutions to NP-hard problems,
in this paper we study and question the practical effectiveness of this approach
for factoring large numbers.
We find no evidence that this is a viable path toward factoring large numbers,
even for scalable fault-tolerant quantum computers,
as well as for various quantum annealing or other special purpose quantum hardware.
\end{abstract}

\ifacm
\maketitle
\fi

\section{Introduction}\label{sec:intro}

In this work we focus on the problem of factoring semi-primes with SAT-solvers.  A
semi-prime $N$ is a composite of two primes $p$ and $q$ which are
roughly of equal size.  These particular composites are conjectured to
be hard to factor, in the sense that no (classical) algorithm or
heuristic is known to factor semi-primes using only polynomially many
resources.  This problem has great relevance for the RSA
cryptosystem~\cite{rsa78}, a widely-deployed public-key cryptosystem.
The RSA cryptosystem is founded upon the difficulty of factoring integers: the
existence of an efficient factoring algorithm would completely break its
security.

Some authors have proposed an alternative approach they refer to as
quantum factoring, and it is occasionally even cited in benchmarks for
factoring~\cite{wikifact}.  In this paper, we explain why these
approaches,
while potentially helpful for studying quantum SAT-solving,
are not likely a viable approach to integer factorization
and, very importantly, are not a meaningful benchmark for people
interested in quantum cryptanalysis of cryptosystems based on the
integer factorization problem.

We attempt to generously extrapolate the kinds of speed-ups one might
expect for a range of quantum solvers, and find no evidence
that this is a viable path toward factoring large numbers, even for
scalable fault-tolerant quantum computers, as well as for various
quantum annealing or other special purpose quantum hardware.

Some researchers only implement quantum factoring for the
purposes of benchmarking the experimental apparatus.  There are several
more relevant algorithms to implement for the purposes of
benchmarking, such as work on randomized benchmarking~\cite{emerson05}
or implementations of quantum error correction.
Framing the experiments as implementations of quantum factoring can
easily be misinterpreted as a meaningful benchmark toward large-scale
integer factorization, and we explain in this article why they are
not.

For many years cryptographers have tracked and benchmarked progress in
classical factorization and attempted extrapolations with an interest in
estimating when RSA schemes with moduli of a given length may be broken using
the number field sieve~\cite{len04, ecrypt18}. The extrapolations take into
account estimates of computing power increase and algorithmic improvements.

This paper highlights why none of the current literature on experimental
implementations of quantum factoring serves the same purpose. In the absence of
a breakthrough that demonstrates factoring can be meaningfully sped up without
a fault-tolerant quantum computer, this sort of tracking of the size of numbers
quantumly factored will only be meaningful after the implementation of several
logical qubits.

One caveat and challenge with tracking and extrapolating is that once
fault-tolerant quantum computers start factoring small numbers,
a constant factor increase in available quantum
resources brings a constant \emph{factor} increase in the size of the number that can be factored
(i.e. we go from being able to factor $n$-bit numbers
  to being able to factor $(cn)$-bit numbers
for some $c>1$ that depends on the factor of increase in time and memory)
because Shor's algorithm runs in polynomial time.
On the other hand, a constant factor increase in classical computing resources only
implies being able to factor numbers that are a few bits
larger using the number field sieve
(i.e. we go from being able to factor $n$-bit numbers
to being able to factor $(n + o(n^{2/3}))$-bit numbers).
Given these quantum scalings,
it will be much harder to reliably extrapolate the size of numbers that can be
quantumly factored, and a relatively small change in computing resources or a
relatively small algorithmic improvement can have a significant impact on the
size of the number that can be quantumly factored.
This is one reason why it is valuable to have post-quantum cryptography
ready for wide-scale deployment before fault-tolerant quantum computers
are available.

The Boolean satisfiability problem (SAT) asks whether there exists an
assignment to the Boolean variables of a given propositional logic
formula such that the formula evaluates to TRUE.\@ This problem was
the first that was proven to be NP-complete~\cite{cook71, levin73}.
Since no algorithms with polynomial runtime for NP-hard problems are
known, solving NP-hard problems has long been considered to be
intractable for real-world computers.  Despite this result, coming
from asymptotic analysis, modern SAT-solvers perform very well on
solving large SAT instances originating from industry and academics,
with formulas that have up to a million clauses~\cite{sat17}.
At the moment of writing there exists no good general method or metric to predict
if a given SAT instance is hard to solve.
For practical applications it therefore makes sense
to assess the performance of the solvers on the investigated instances
by careful benchmarking instead of doing asymptotic analysis.

The original goal of this project was to encode the RSA factoring
challenges~\cite{rsafact} to SAT instances and see how well modern SAT
solvers would perform on those instances.  The smallest semi-prime of
these challenges is RSA-100: a 100-digit or 330-bit number.  This
number was factored in a few days almost immediately after the
challenge was posted~\cite{dl94} in 1991, whereas the current record
for factoring stands at factoring RSA-768: a 768-bit
semi-prime~\cite{rsa768}.  The intention was to compare current
state-of-the-art SAT solvers against the numerical results from 1991,
but it turns out that even the smallest RSA semi-prime poses too big
of a challenge for these solvers.

\subsection{Contributions}\label{ssec:contributions}

This work provides a numerical analysis on the hardness of factoring
numbers by solving the corresponding satisfiability problem, thereby
confirming the folklore that factoring numbers does indeed give
``hard'' SAT instances.  This is done by measuring the speed of the
currently fastest SAT solver.  We justify the choice of numerical
analysis over theoretical asymptotic analysis by applying some common analysis tools
from modern SAT solving theory and the observation that the tools provide
no good prediction for the actual runtime.
We extrapolate the numerical results to investigate the asymptotic behavior of the
solver and compare the results with the asymptotics of factoring with
numerical algorithms.  Finally, the results are used to estimate an upper
bound on the speedup that can be achieved on this specific problem
using currently known quantum algorithms.

As a minor contribution, we developed a tool that can create smaller
SAT instances for factoring\footnote{using long multiplication} than
any other publicly available tool.  This tool and scripts for
generating semi-primes and reproducing the results of this paper have
been made available online~\citerepo.

\section{SAT instances}\label{sec:instances}

An instance of the SAT problem is a formula in Boolean propositional
logic: every \emph{variable} ($x$) can take the value TRUE or FALSE as
specified by the respective \emph{literals} $x$ and $\bar{x}$.  This
work considers the equivalent~\cite{karp72} CNF-SAT where all formulas
are in conjugate normal form (CNF): each formula must be a conjunction
of disjunctions of literals.\footnote{Further restricting each clause
  to exactly three literals would give the equivalent 3SAT problem.}
The disjunctions are often called \emph{clauses}.  A satisfying
assignment gives a value to each variable such that at least one
literal evaluates to TRUE in every clause.  All tools we used for
generating and solving SAT instances work with the DIMACS
format which specifies formulas in CNF form.

Another (equally hard) formulation of the problem is called
CircuitSAT:\@ given a Boolean circuit with a single output, is there
an input such that the output is TRUE?\@  One can translate any Boolean
circuit into a Boolean formula: assign a variable to each wire and
let the clauses describe the gates.  For example the Turing complete
NAND-gate with input wires $x$, $y$ and output wire $z$ has the
corresponding formula $(x \lor z) \land (y \lor z) \land (\bar{x} \lor
\bar{y} \lor \bar{z})$.  Simulating gate execution is done by fixing a
value on the input wires: for example by adding the clauses $x \land
\bar{y}$.  A SAT-solver can examine those five clauses and find that
the only satisfying assignment sets $z = \mbox{TRUE}$.  Combining
gates to make a circuit is done by reusing output variables of earlier
gates as input variables in later gates.

More interesting is to fix a value on the output variables of a
circuit and ask the SAT-solver to find a satisfying assignment.  For
example adding the clause $z$ to the NAND-gate gives three satisfying
assignments: $x \land \bar{y}$, $\bar{x} \land y$, and $\bar{x} \land
\bar{y}$.  In general a circuit might have zero or more satisfying
assignments.  Effectively the SAT-solver is finding preimages to the
function described by the circuit.
An immediate cryptanalytic application that springs to mind is finding
preimages to secure hash functions: indeed this has been done with
varying results~\cite{mz06,ms13,dkmnpw17}.  More general cryptanalytic
applications can be found throughout literature~\cite{mm00} and occur
in modern benchmarks~\cite{sat17}, although asymmetrical cryptographic
primitives are rarely targeted.

This work examines circuits that encode the multiplication of two
integers $p$ and $q$.  We fix the multiplication output bits of the
circuit to the bit-values of the semi-prime $N$ and ask the SAT-solver
to find a satisfying assignment.  Only two exist\footnote{The specific
  encoding ensures the trivial solutions $N=1N$ and $N=N1$ do not give
  satisfying assignments.}: those representing $N=pq$ and $N=qp$, so
from the assignment one can read the factorization of $N$.  For the
remainder of this paper $n$ represents the size of $N$ in bits.  We
limit $p$ and $q$ similar to how the RSA cryptosystem limits its
parameters: both need to be equally sized primes. We interpreted this
last requirement to mean that their most significant bit may differ
by at most one position.

\subsection{Encoding}\label{ssec:encoding}

Despite the asymptotic worst-case exponential runtime associated with SAT
instances, it turns out to be non-trivial to generate ``hard'' SAT
instances: instances where the solver runtime grows exponentially in
the number of variables.  For many instantiations of the SAT problem,
it turns out that the average case can be solved relatively efficient
with modern SAT solvers.  Specialized tools such as
ToughSat~\cite{toughsat} exist that can generate SAT instances that
are hard on average, based on problems such as integer factorization.

Multiplying larger integers requires larger circuits, which leads to
instances with more variables and clauses, which leads to longer
solving times.  However, there are many choices to make when computing
multiplication in a circuit and each choice will lead to different
encodings of the SAT instance and a different solver runtime.
For SAT solvers in general it turns
out that the details of the encoding of a problem (beyond metrics such
as number of variables and clauses) can have a significant impact on
the solver runtime.
The first choice is to consider different multiplication algorithms: a simple
one and a more complex encoding that in theory leads to smaller instances.

Long multiplication (or schoolbook multiplication) is computed by
multiplying $p$ by each digit (bit) of $q$ and adding the shifted
results.  For multiplying two $m$-bit numbers (where $m=n/2$) this
requires $\Theta(m^2)$ bitwise multiplications and additions.
The exact number of operations depends mainly on the circuit
used for addition: our tool for generating instances~\citerepo{}
minimizes the number of both variables and clauses by maximizing
the number of full-adders used in the circuit.
Counting the variables in the generated instances and
applying regression reveals that the
number of variables grows approximately as $0.750n^2 + 0.496n - 2.05$ and
similarly the
number of clauses grows as $4.25n^2 - 4.01n - 9.87$ with on average
3.31 literals per clause.

Karatsuba multiplication~\cite{kar63} asymptotically improves upon long
multiplication by a divide-and-conquer strategy and requires only
$\Theta(m^{\log_2 3})$ multiplications at the cost of requiring more
additions.  The instances we tested were generated by the ToughSat
application~\cite{toughsat} and contain approximately $2.59n^{\log_2 3} -
7.57n + 8.75$ variables and $61.5n^{\log_2 3} - 170n - 386$ clauses
with on average 6.77 literals per clause.  Inspection of the generated
instances reveals that the Karatsuba circuits were built
from more complex gates, which explains why there are more literals
per clause.  It is likely that building the Karatsuba
circuit with a similar gate set would increase the number of variables
and clauses by another (constant) factor.

Asymptotically the Karatsuba algorithm is not the best known algorithm
and is outperformed by for example Toom-Cook or FFT-multiplication.
Given that these methods introduce additional overhead for small instances and
given the minor difference in the runtime of long multiplication and
Karatsuba (see \hyperref[sec:classical]{Section~\ref*{sec:classical}}), 
it appears that the cross-over point where these algorithms are faster
vastly exceeds a feasible instance.

Hardware design provides alternative multiplication algorithms, which
are often optimized to minimize latency and for various other physical
constraints.  There is no indication that these optimizations are
related to optimizations that lead to smaller and/or easier SAT
instances.  In fact our adder encoded in the SAT instances minimizes
the number of half-adders required, which gives the smallest number of
variables and clauses and results in the fastest SAT solver times, but the
resulting clauses encode a circuit that would give extremely high
latency if built from physical components.

Since the multiplication circuit is the same for each
semi-prime of the same bitlength there is an alternative strategy we
can apply when we want to factor only one of many semi-primes.
We encode the multiplication circuit once and then
``fanout'' the resulting wires to circuits that check if the output
equals a semi-prime.  Those results are combined with a large
OR-gate, so that the entire instance evaluates to TRUE if the multiplication
outcome is equal to any of the semi-primes.  By inspecting which values were
assigned on the circuit input wires by the solver we learn which of the semi-primes
it actually factored.
The idea behind this encoding is
that if there is an easy semi-prime somewhere in the input, then the solver
itself may detect this and focus on solving that instance.  As long as we
encode only polynomially many semi-primes in the instance, the total instance
size will remain polynomial.

An alternative solution for factoring numbers with SAT is to encode
the integer division circuit $N / p = q + r$ and fixing the
input value $N$ and output remainder $r=0$.  The rationale for this
encoding is that the solver would only have to assign values to the
bits of $p$ and can then deterministically evaluate the entire circuit
and check if the remainder is zero.  However, in practice this
encoding leads to substantially larger SAT-instances and tests with
various solvers indicate that solving such instances is significantly
slower, so we did not investigate this encoding any further.


A more promising approach is to reduce some subroutine of the NFS to SAT where
there is little or no increase in complexity by mapping to SAT, analogous to
the approach taken in~\cite{bbm17}.  In this case, even a small quantum speed-up will
lead to a faster integer factorization algorithm.  This approach is studied in
detail in~\cite{mvv19}.

\section{Classical Solvers}\label{sec:classical}

Modern SAT solvers come in two classes.  Conflict-Driven Clause
Learning (CDCL)~\cite{dp60, dll62} combines conflict analysis with
branch heuristics to systematically backtrack the search-space of an
instance.  Stochastic local search approaches such as employed by
WalkSAT~\cite{walksat} or simulated annealing combine randomized
assignments with probabilistic updates to find assignments that
minimize the number of clauses violated.
We found that for the semi-prime instances CDCL solvers outperformed
the local search solvers by an order of magnitude.
The scope of this project is limited to the black-box analysis of
publicly available SAT solvers.  This means we will not investigate the
internals of the solvers for analysis of the runtime, nor do we allow
domain-specific knowledge to speed up solver times.

We tested the MapleCOMSPS~\cite{maplecomsps} SAT solver for the simple
reason that at the time of running the benchmarks this was the fastest solver
according to the SAT Competition 2016~\cite{satcomp16}.
We compiled and ran the solver with
default settings, except for the random seed which was fixed for each
call to the solver to ensure
reproducibility of the results.

Another solver that we tested is CryptoMiniSat 5~\cite{cryptominisat5},
because it has ``Automatic detection of cryptographic
[\dots{}] instances''~\cite{cryptominisat2}.  One might consider this
to be cheating by using domain-specific knowledge and therefore it
should not be included in the benchmarks.
CryptoMiniSat appears to focus on symmetric cryptography and appears
to provide no speedup on public cryptography instances, which we
confirmed during an initial
round of benchmarking. We inspected the (partial) results and found that
CryptoMiniSat 5 was consistently being outperformed by MapleCOMSPS.\@
For this reason we did not further analyze this solver, but the results
can be found in \autoref{app:cms5}.

All measurements were performed on a ThinkPad laptop with a 64-bit
Intel Core i5--4200M (Haswell) CPU running at 2.50GHz.  All
measurements were executed sequentially and on a single core.  Where
applicable we use
regression to fit a line to the data and the
goodness-of-fit is quantified by the $r^2$ parameter.

\subsection{Results}

\ifacm
\begin{figure}
  \includegraphics[width=\columnwidth]{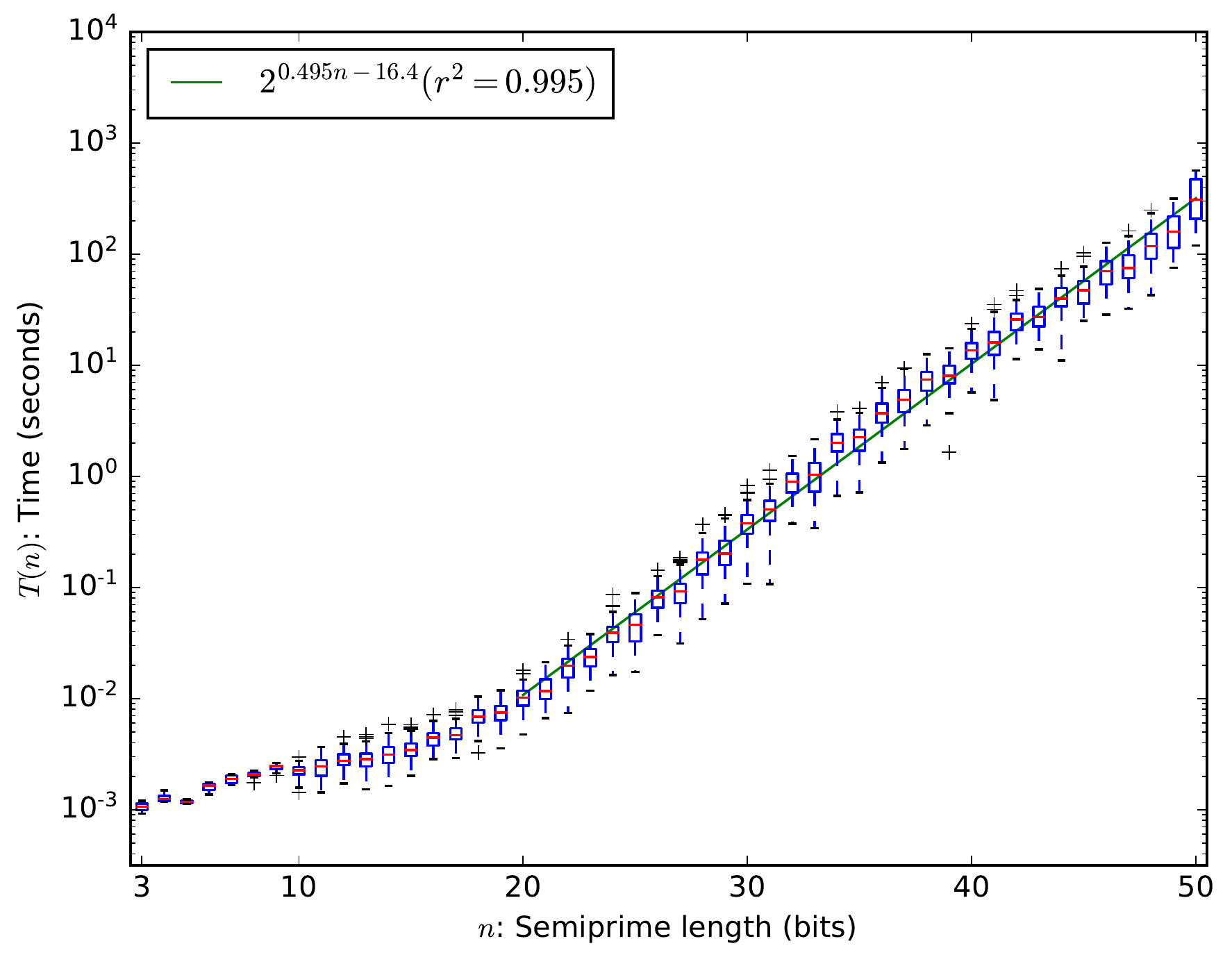}
  \caption{Runtime of factoring with MapleCOMSPS: schoolbook multiplication}
  \label{fig:maplelong}
\end{figure}
\begin{figure}
  \includegraphics[width=\columnwidth]{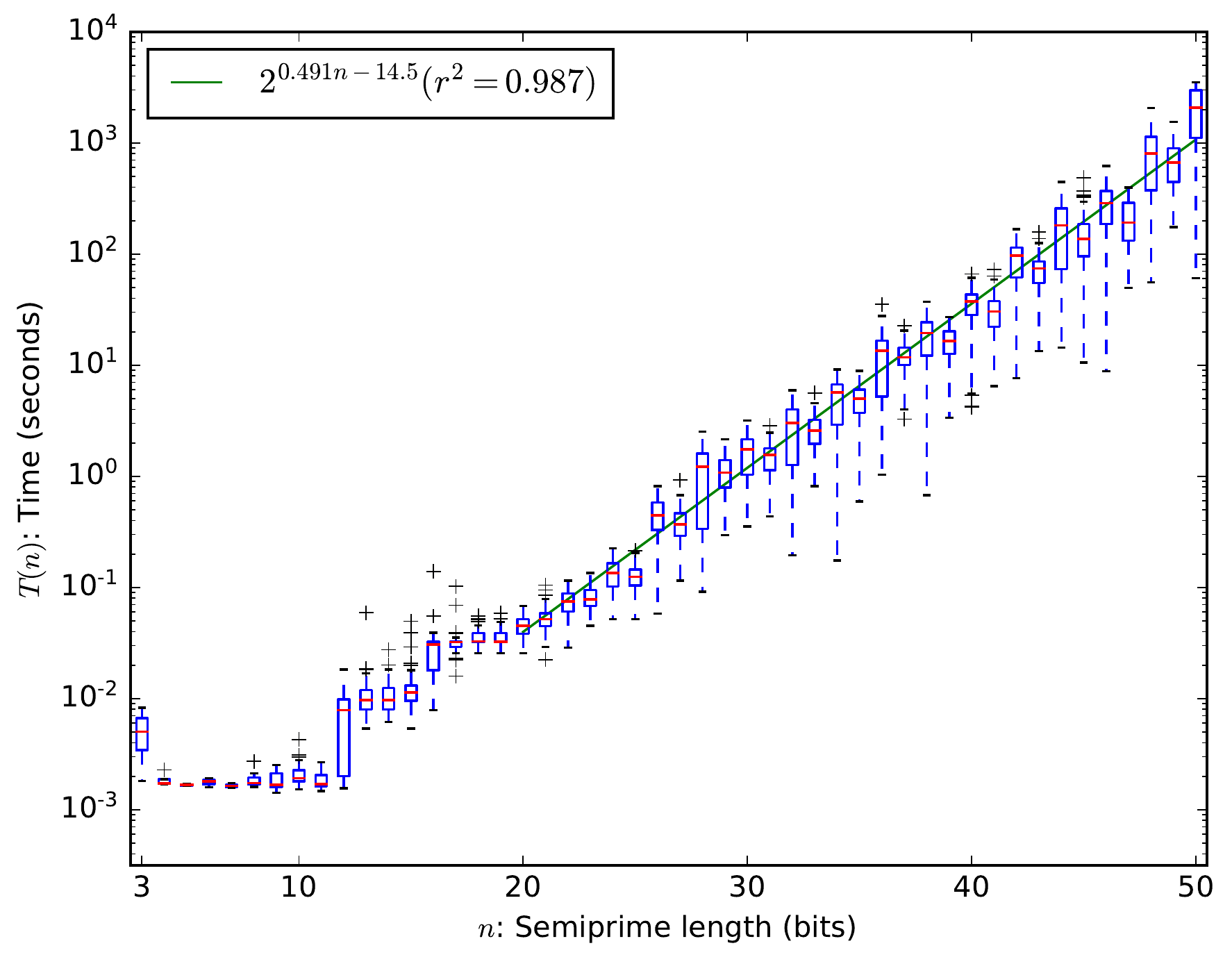}
  \caption{Runtime of factoring with MapleCOMSPS: Karatsuba multiplication}
  \label{fig:maplekar}
\end{figure}
\else 
\begin{figure}[ht!]
  \centering
  \begin{subfigure}[t]{.49\textwidth}
  \includegraphics[width=\textwidth]{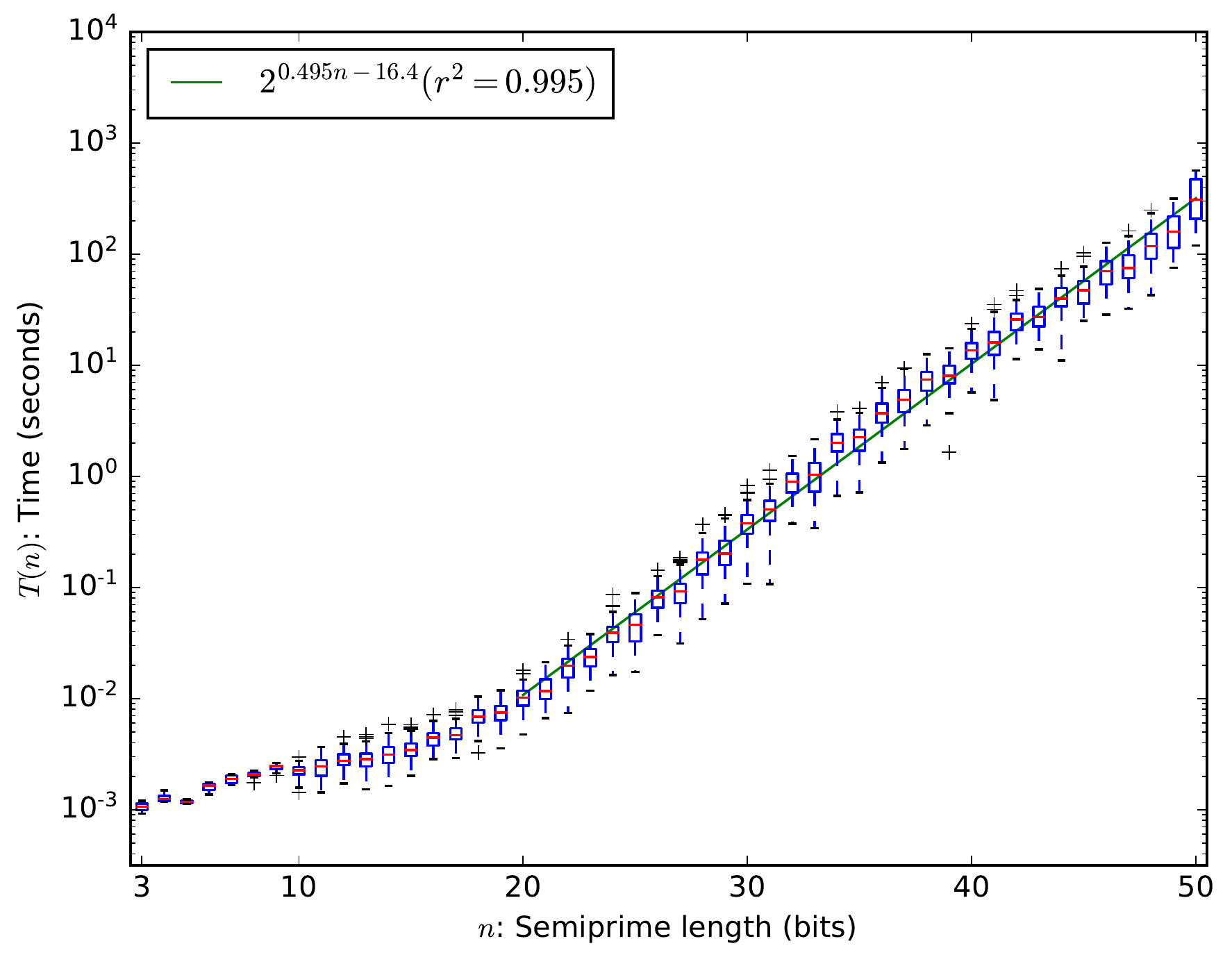}
  \caption{schoolbook multiplication}
  \label{fig:maplelong}
  \end{subfigure}
  ~
  \begin{subfigure}[t]{.49\textwidth}
  \includegraphics[width=\textwidth]{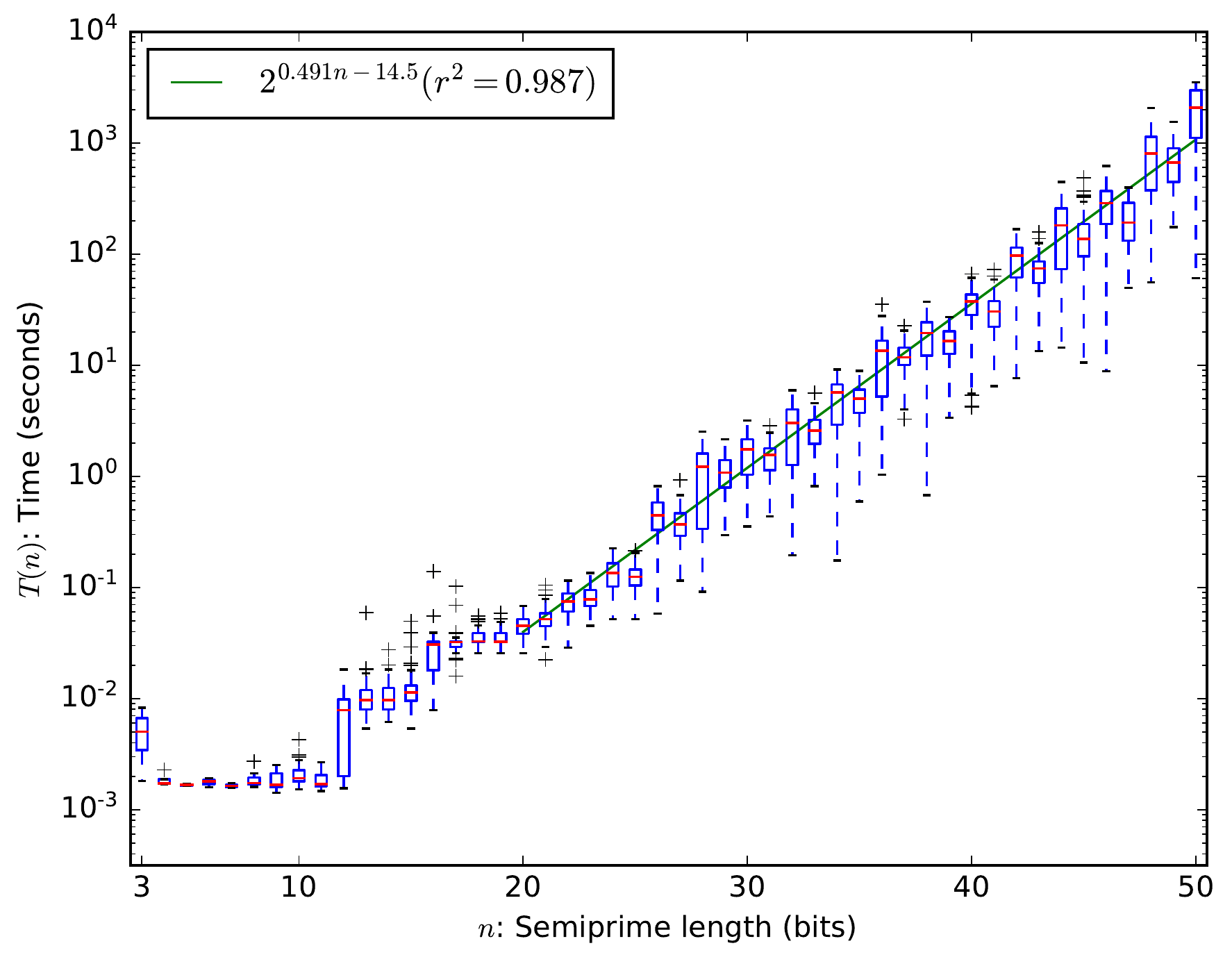}
  \caption{Karatsuba multiplication}
  \label{fig:maplekar}
  \end{subfigure}
  \caption{Runtime of MapleCOMSPS on factoring semi-primes.}
  \label{fig:maplemean}
\end{figure}
\fi 

Usually when analyzing the runtime of a randomized algorithm we are interested
in the expected runtime: the mean computed over the random bits.  We do this by
factoring the same number multiple times using a different PRNG-seed for the
solver and average the runtime to compute the expected runtime numerically.
We are interested in the asymptotics: the growth of the runtime as a function
of the size of its input, so we group the semi-primes by their bitlength $n$
(100 semi-primes per bitlength) and plot the mean runtime of solving five times.
The results are
given in \autoref{fig:maplelong} and are showing an exponential trend.
The green line is fitted against the median runtime of all semi-primes of the
same bitlength.
 
We repeated the same experiment for multiplication with the Karatsuba algorithm.
The results are given in \autoref{fig:maplekar}:
note that asymptotic runtime has improved somewhat over schoolbook
multiplication at the cost of a larger constant. 
We conclude that changing the multiplication algorithm does not make factoring
with SAT solvers efficient.
Since the larger constant dominates the runtime at this small scale, we will
consider schoolbook multiplication for the remainder of our experiments.

\ifacm
\begin{figure}
  \includegraphics[width=\columnwidth]{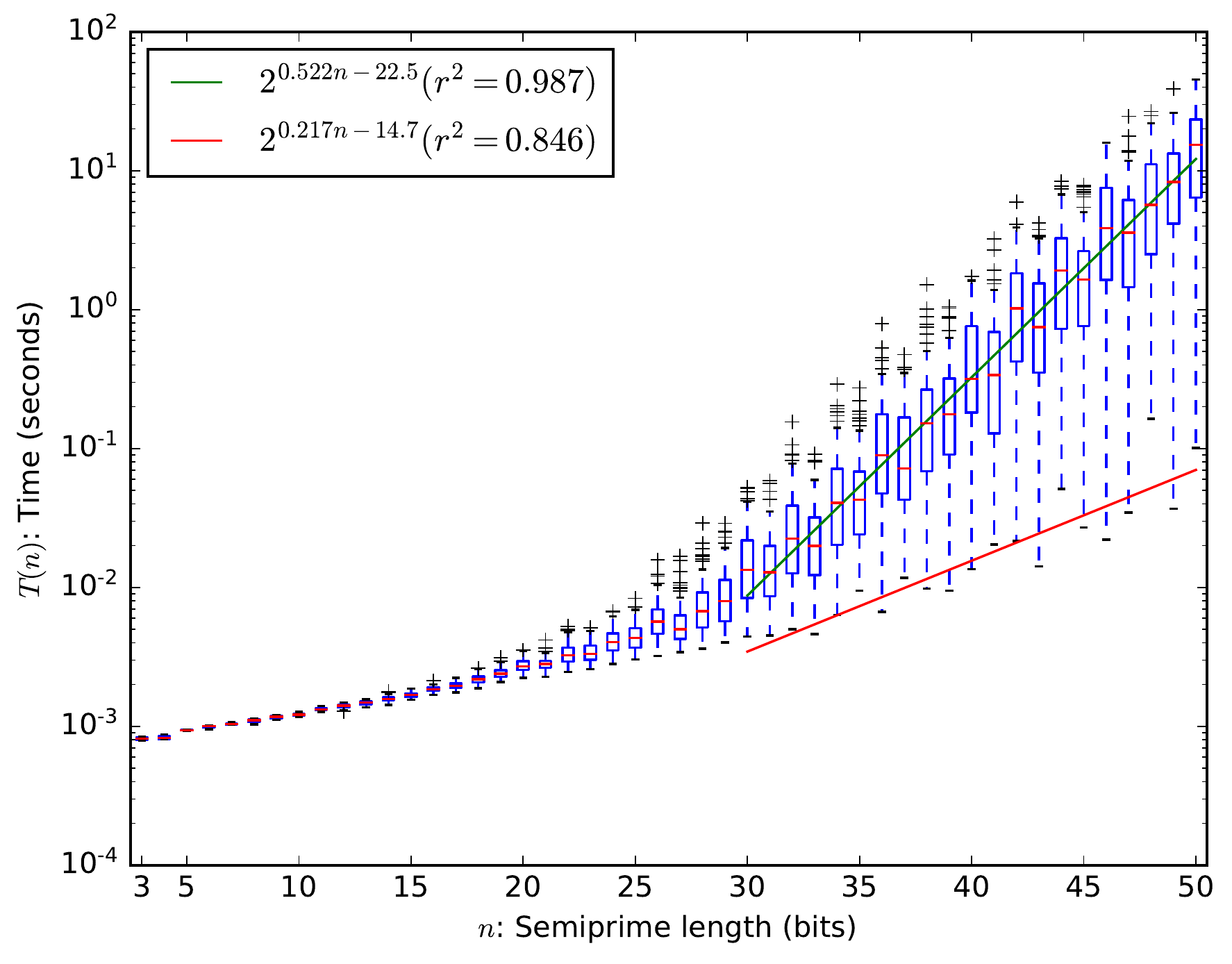}
  \caption{Minimum runtime of MapleCOMSPS on factoring semi-primes
  using schoolbook multiplication.}
  \label{fig:maplelongmin}
\end{figure}
\else 
\begin{figure}[ht!]
  \centering
  \includegraphics[width=.6\textwidth]{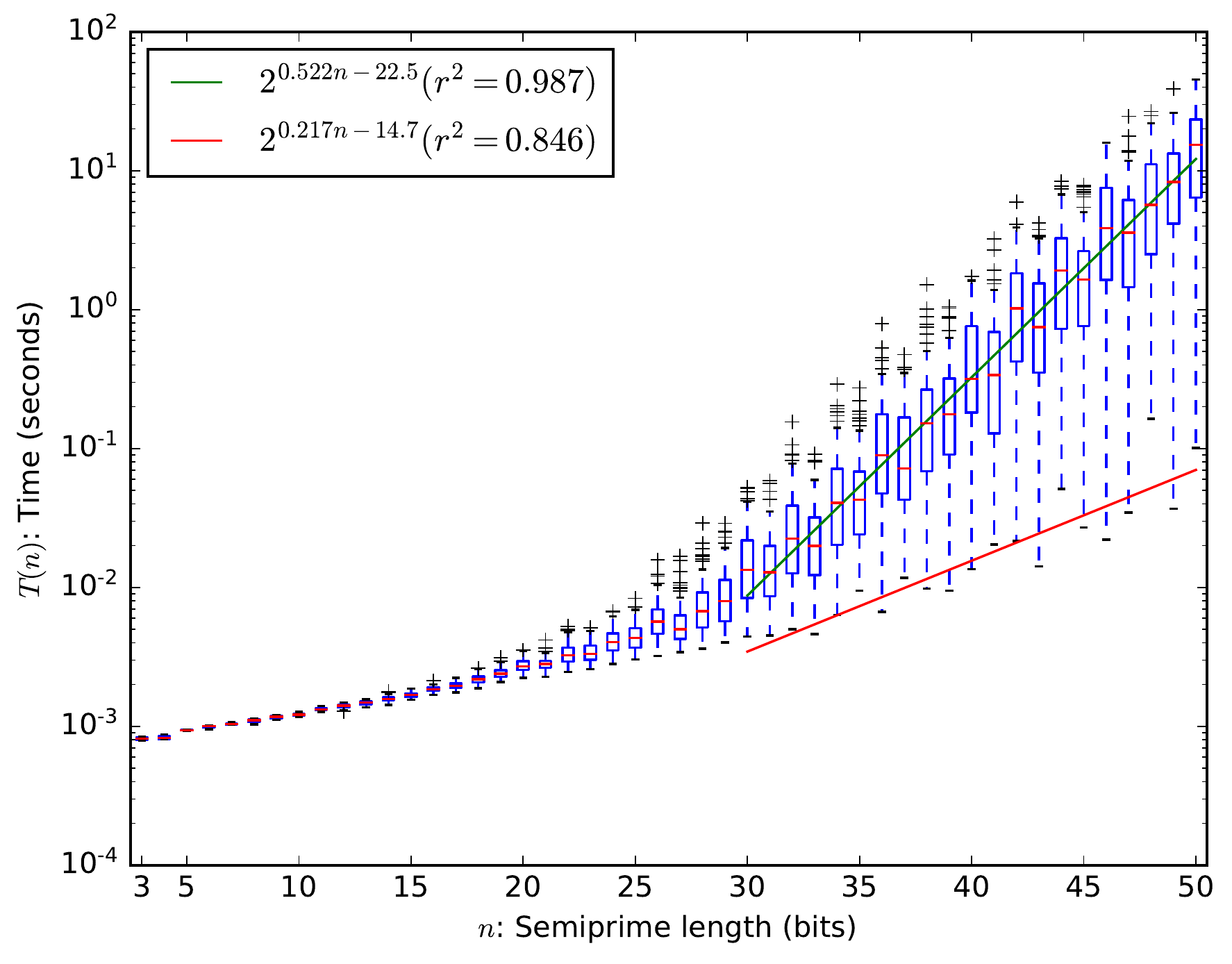}
  \caption{Minimum runtime of MapleCOMSPS on factoring semi-primes
  using schoolbook multiplication.}
  \label{fig:maplelongmin}
\end{figure}
\fi 

An alternative strategy for factoring is to run several solvers in parallel and
wait for the first one to return a solution. 
We simulate this strategy by taking the minimum solver time of solving the
same instance with the solver initialized with 100 different random seeds for
100 semi-primes per bitlength:
the results are given in \autoref{fig:maplelongmin}.
Asymptotically the runtime became worse by employing this strategy.
Note that this strategy does push down the constant by approximately $2^{6.1}$.
Since this is smaller than 100 it does not
lead to a lower expected runtime on this small scale when we consider the
total runtime of all parallel solvers.

We can also see in \autoref{fig:maplelongmin} that some semi-primes are significantly
easier to solve than others with this strategy.  Even if we only
manage to factor some semi-primes that may be important to (for example)
cryptography.
For this method to be asymptotically efficient, it is required that the
runtime is pushed down exponentially for more than just negligibly many cases.
To see if it does we can inspect the
distribution of the solver runtime given different seeds. Here we focus on three
different semi-primes\footnote{the distribution for all other semi-primes
can be generated at~\citerepo{}}:
the easiest, average and hardest semi-prime from the
100 semi-primes of 35 bits, where hardness is defined by the expected (mean) solve
time computed over 360 seeds.

\ifacm
\begin{figure}
  \includegraphics[width=\columnwidth]{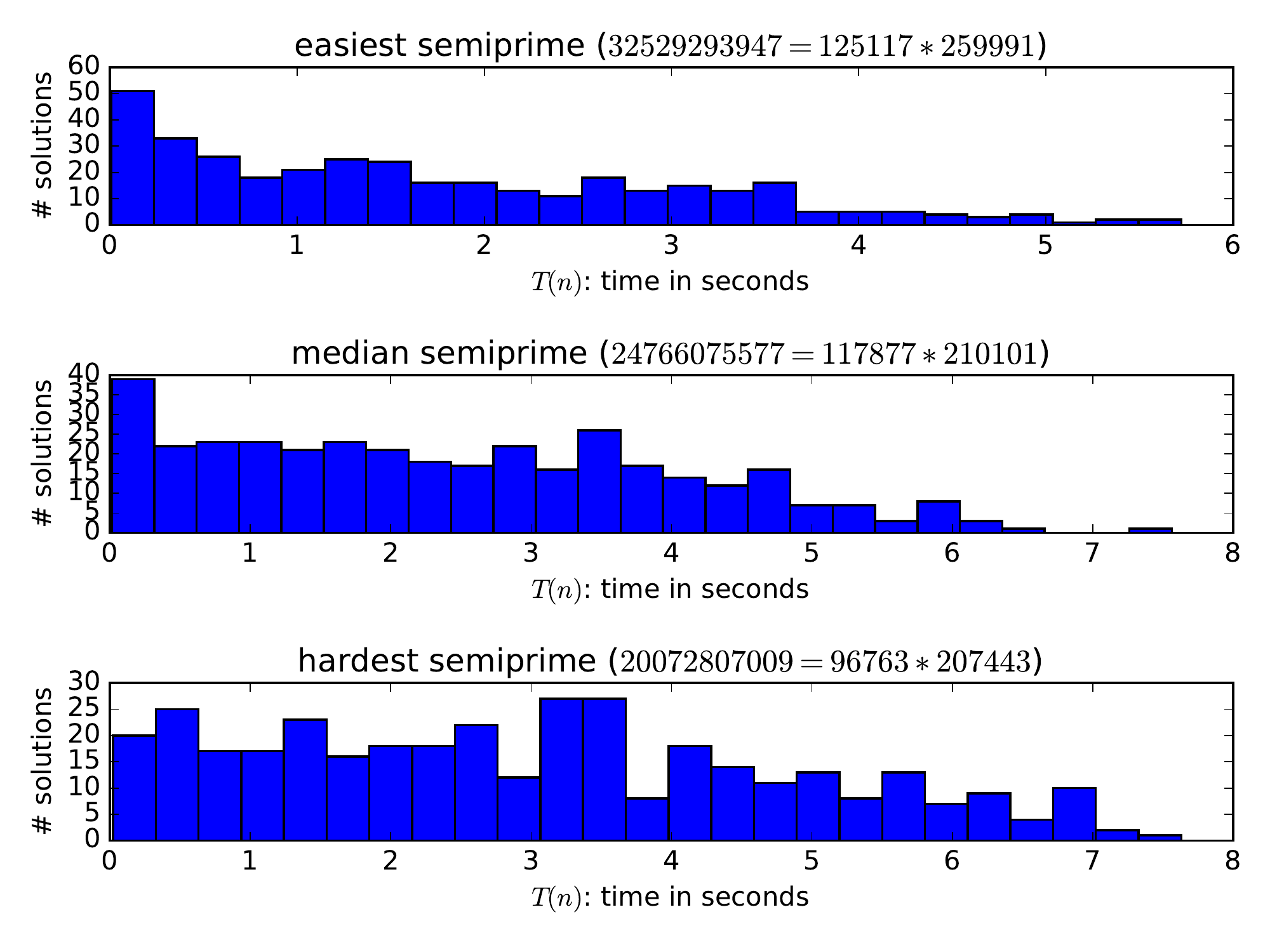}
  \caption{Histogram of MapleCOMSPS runtime (linear time-scale)}
  \label{sfig:zoomed35}
\end{figure}
\begin{figure}
  \includegraphics[width=\columnwidth]{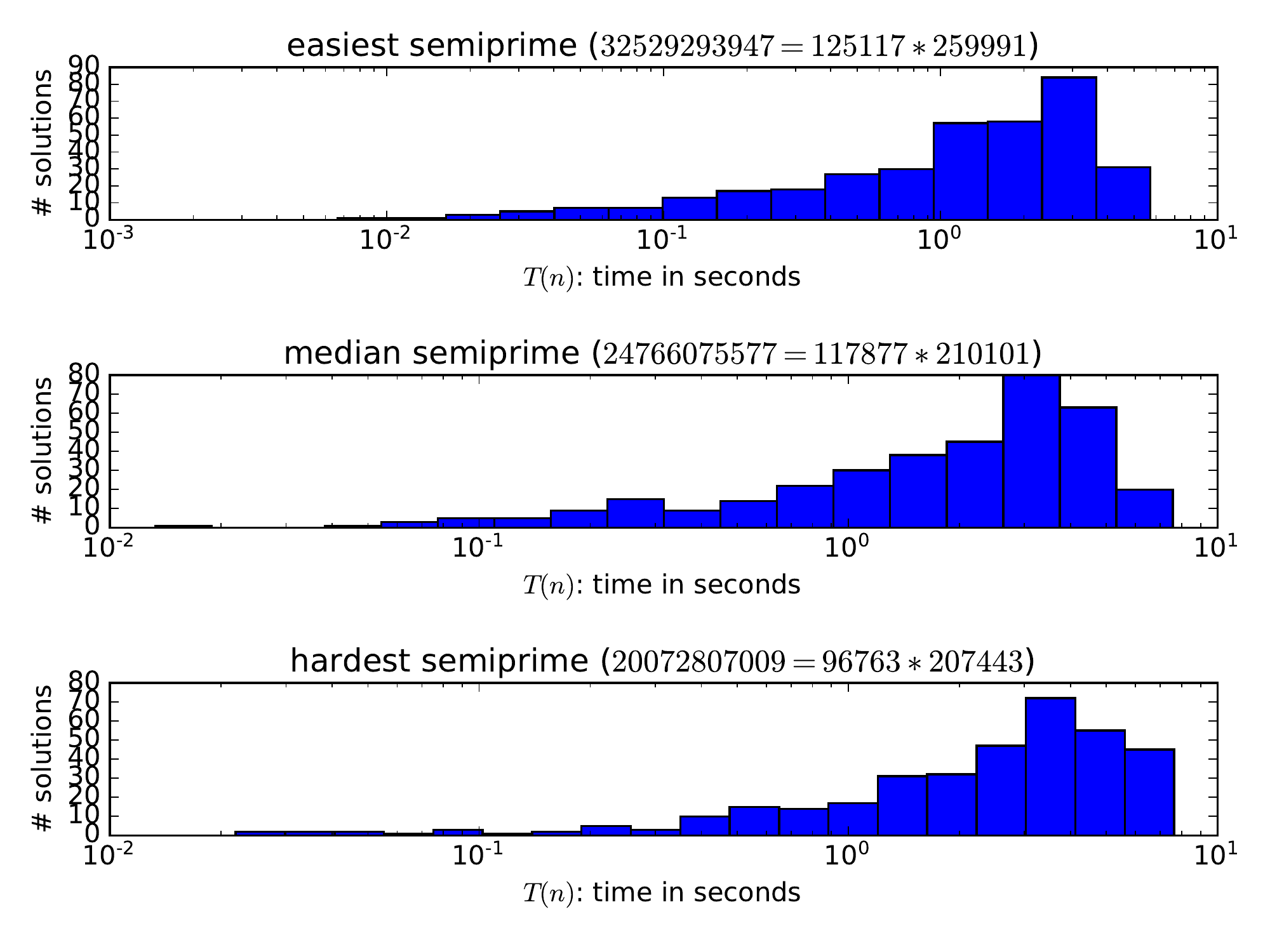}
  \caption{Histogram of MapleCOMSPS runtime (logarithmic time-scale)}
  \label{sfig:zoomedlog35}
\end{figure}
\else 
\begin{figure}[ht!]
  \centering
    \begin{subfigure}[t]{.49\textwidth}
      \includegraphics[width=\textwidth]{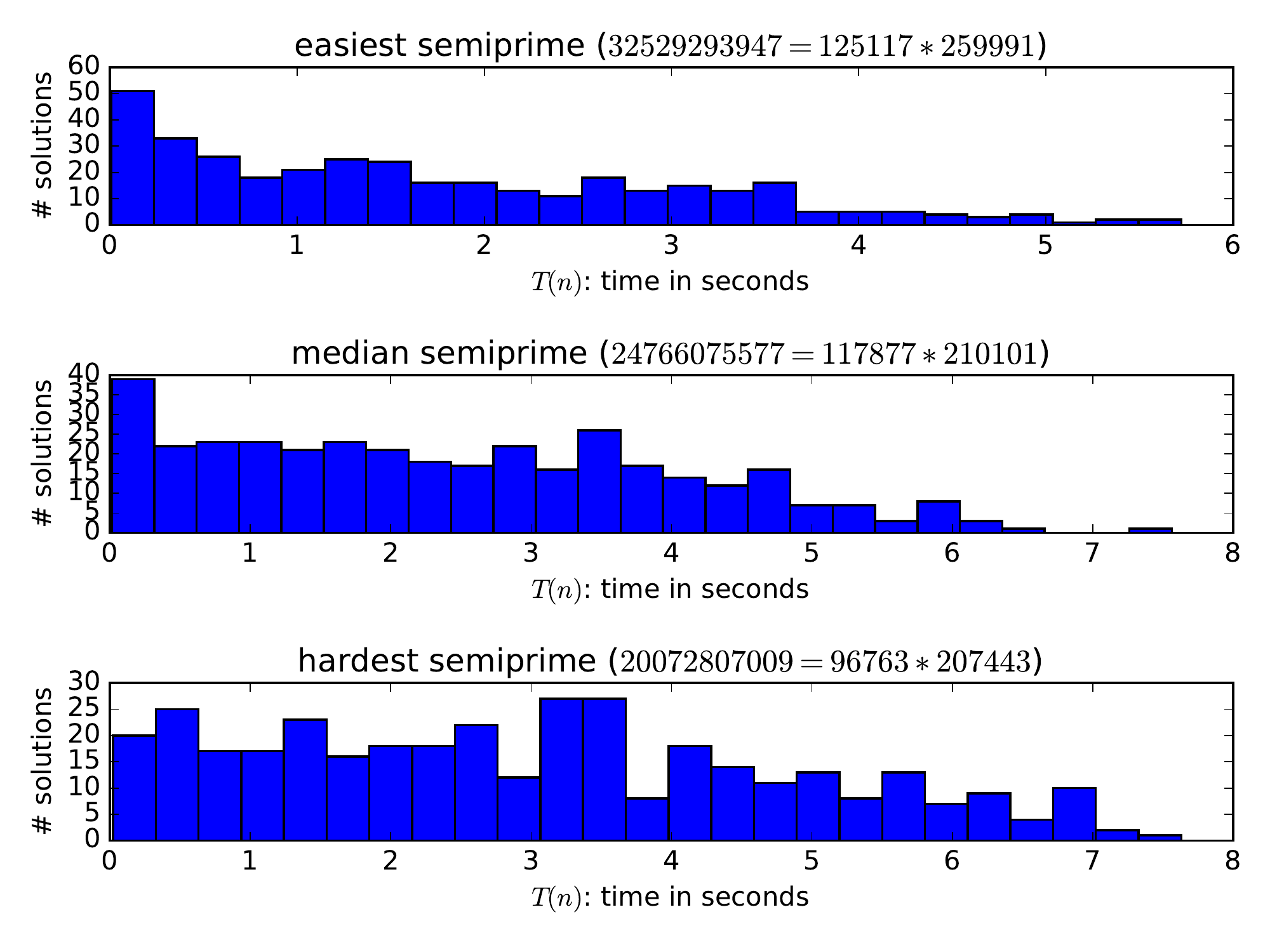}
        \caption{Linear time-scale}\label{sfig:zoomed35}
    \end{subfigure}
    ~
    \begin{subfigure}[t]{.49\textwidth}
      \includegraphics[width=\textwidth]{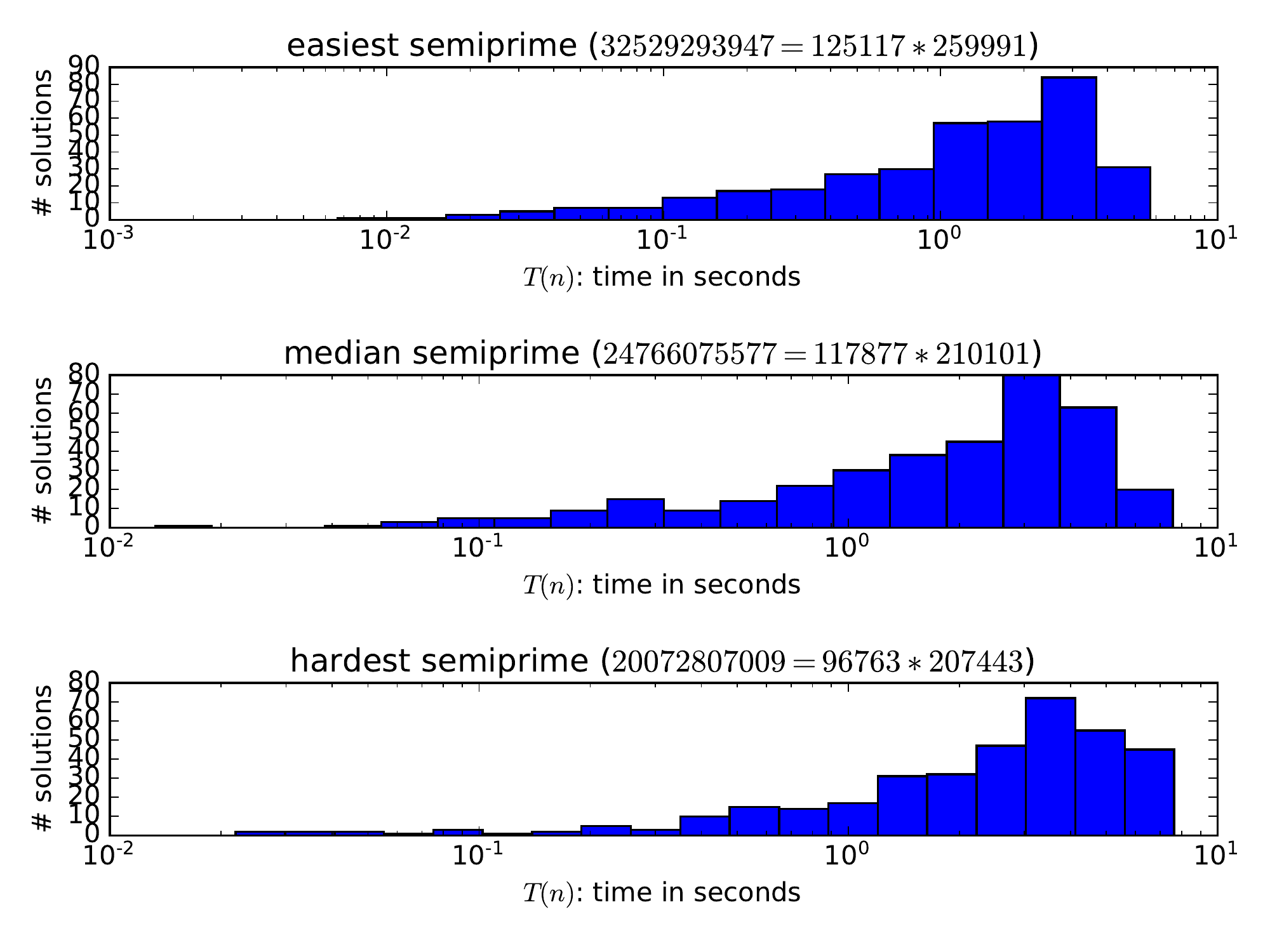}
        \caption{Logarithmic time-scale}\label{sfig:zoomedlog35}
    \end{subfigure}
    \caption{Histogram of the MapleCOMSPS runtime on factoring semi-primes using schoolbook multiplication.}
  \label{fig:hist}
\end{figure}
\fi 

Although no strong conclusions should be drawn from the results in \autoref{sfig:zoomed35}, the
distribution does suggest that running a few parallel solvers may
lower the total runtime.  To see if it may be considered efficient we again
inspect the distribution but this time on a logarithmic scale: see \autoref{sfig:zoomedlog35}.

This data suggests that even if the method could push down the runtime significantly
for any semi-prime, it only does so with negligible probability.
Another way of interpreting this data is that employing parallel SAT solvers to factor a
semi-prime does not appear to be better than employing a single solver.

\ifacm
\begin{figure}
  \includegraphics[width=\columnwidth]{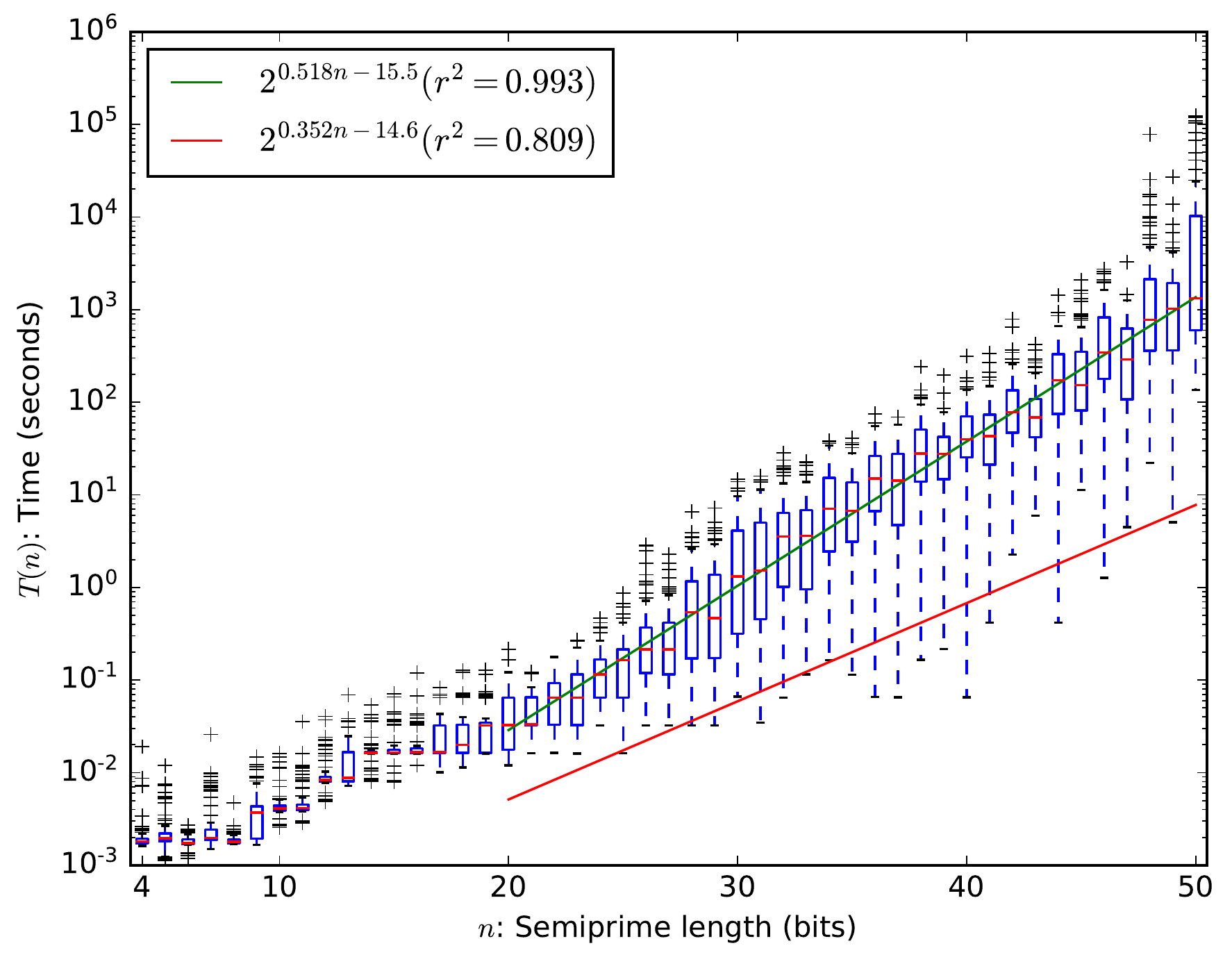}
  \caption{MapleCOMSPS runtime factoring one out of a hundred semi-primes per instance.}
  \label{fig:zoomedlog35}
\end{figure}
\else 
\begin{figure}[ht!]
  \centering
  \includegraphics[width=.6\textwidth]{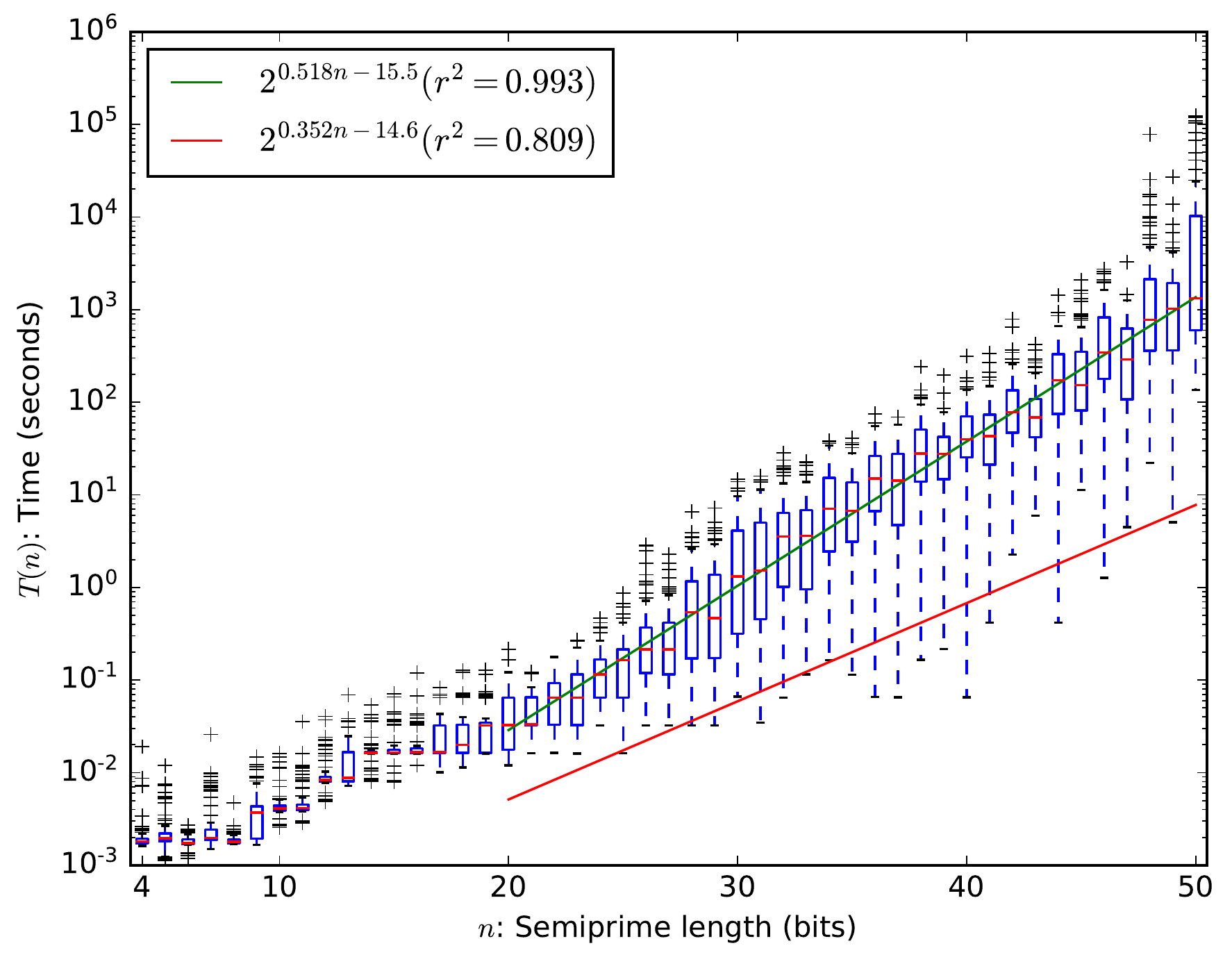}
    \caption{Runtime of MapleCOMSPS on factoring one of 100 semi-primes encoded in each
    instance using schoolbook multiplication.}
  \label{fig:zoomedlog35}
\end{figure}
\fi 

The last strategy we investigate is that of encoding multiple semi-primes
into a single instance.  We encoded 100 semi-primes per bitlength in each
instance and solved it 100 times using different seeds.  The results are
given in \autoref{fig:zoomedlog35} Note that whereas
the vertical boxplots in previous plots show a distribution over different
primes, here a distribution over different solver PRNG-seeds is shown.
From the data we conclude that this strategy is less efficient than solving
instances with a single semi-prime.
From inspection of the solver solution we can see
which semi-prime was factored (see~\citerepo{}).
This reveals that some
semi-primes in the same instance are factored more often than others,
suggesting that these are easier to factor by the solver, although we
note that these are not ``easy enough'' to make the overall method efficient.

\subsection{Patterns}

The above results support that it is hard \emph{on average} to
factor a number with SAT solvers, but we can also observe that some numbers are easier to factor than
others.  It is an interesting question whether there is some structure
in the semi-primes that is picked up by solver that allows it to
factor more efficiently or whether the solver's heuristic choices
accidentally lead to a faster solution.  We try to answer this question
by inspecting the instances using two analytic methods from the SAT
literature (backdoors and community structure) and we do some
manual inspection of the instances.  Because we are interested in the
fastest solver time, we focus on the minimum solver time per instance
given different random seeds.

\subsubsection{Backdoors}

Backdoors in SAT instances were introduced by Williams, Gomes and
Selman~\cite{wgs03}.  A backdoor is a subset of variables such that
setting these variables to any value allows a so-called subsolver to
assert if the entire formula is satisfiable in polynomial time.  If
the solver can find such a backdoor of size $k$ with a subsolver that
runs in time $l$, the entire solver can run in time $O(l2^k)$.  Any
CircuitSAT instance has a trivial backdoor in the form of the input
wires/variables: set these and the rest of the clauses can be
determined deterministically.\footnote{This is why we also encoded a
division circuit: the input to that circuit contains 
one prime instead of two.}
A backdoor subset for CircuitSAT therefore only becomes interesting when it is
smaller than the set describing the input variables.

Every instance has $n$ input wires, but the solver runtime suggests
that a backdoor of $k \approx n/2$ variables was found.  Given the
structure of the problem this is not surprising (division to find the other
$k$ input bits only takes polynomial time), but it is somewhat
surprising given that it is unlikely that the SAT solver was programmed
to perform this division.
More meaningful analysis of this observation would require inspecting the internals of the solver to look for
potential subsolvers and backdoor detection capabilities.  We consider
this outside the scope of this project.  We simply conclude that even
if a backdoor of size $n/2$ is found then the runtime of the SAT
solver would still be exponential and therefore would not
impact the security of RSA.

\subsubsection{Community structure}

A SAT instance can be represented as a graph where each variable is a
vertex and an edge is drawn between vertices when the variables occur
in the same graph.  The community structure of a graph is often
characterized by a quality metric $Q$ (also known as the modularity of
the graph).  According to Newsham, Ganesh, Fischmeister, Audemard and
Simon~\cite{ngfas14} the community structure of a SAT instance should
provide us with a good prediction on how hard it is to solve that
instance: instances are harder to solve when $0.05 \leq Q \leq 0.12$.

An immediate problem that occurs when applying the above theory to the
generated circuits is that all instances for semi-primes of the same
bitlength have the same structure: the encoded circuit is simply an $m$ by
$m$ bit multiplier.  Therefore, we compute the community structure
only on the instances after they are simplified by the solver's
preprocessor.  We approximate the value of $Q$ with the
greedy algorithm by Clauset, Newman and Moore~\cite{cnm04}.

\ifacm
\begin{figure}
  \includegraphics[width=\columnwidth]{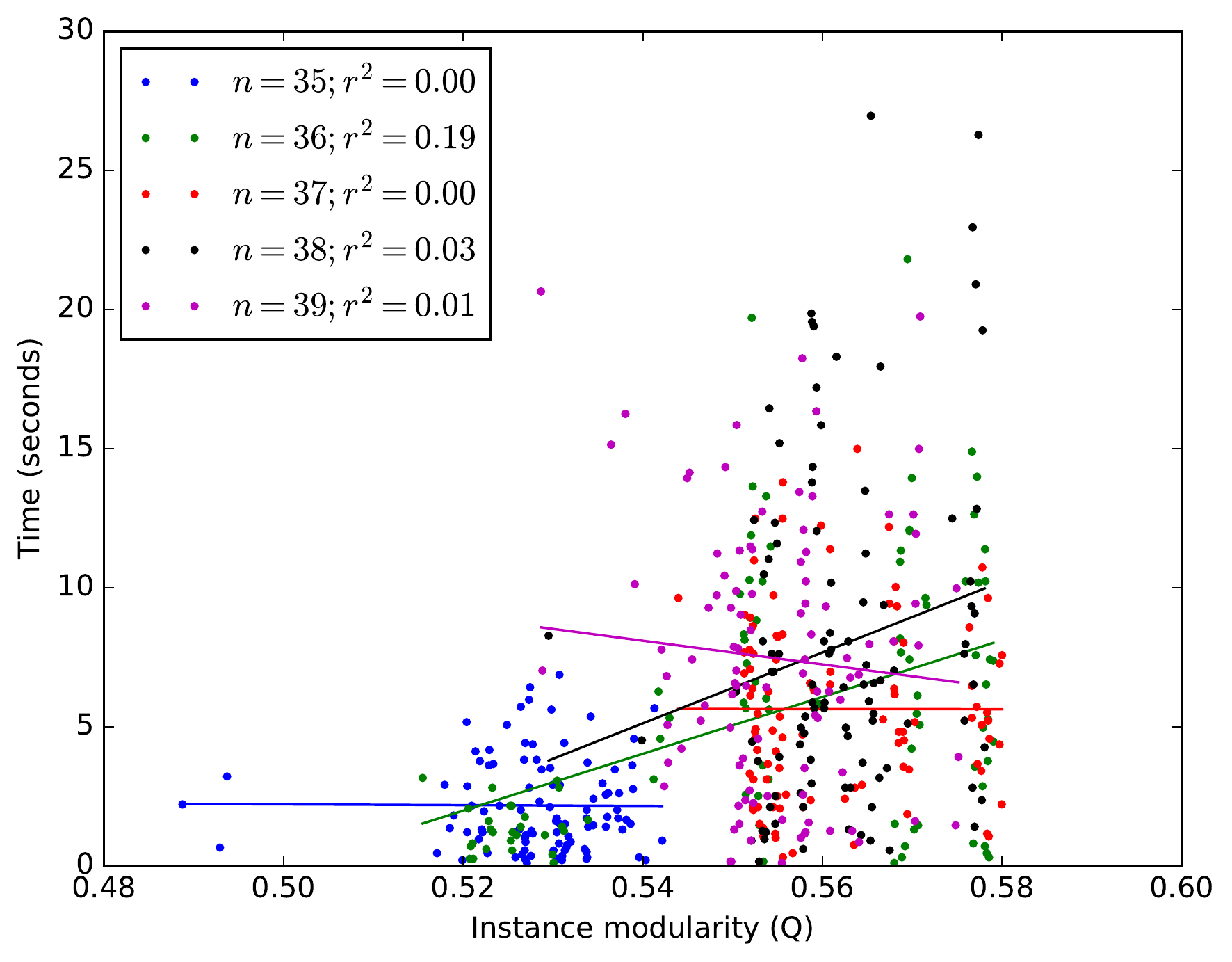}
  \caption{Solver times per community structure.  We only display the results
    for some values of $n$ to avoid more clutter, but similar results
    hold for all $n$.}
  \label{fig:commstruct}
\end{figure}
\else 
\begin{figure}[ht!]
  \centering
  \includegraphics[width=.49\textwidth]{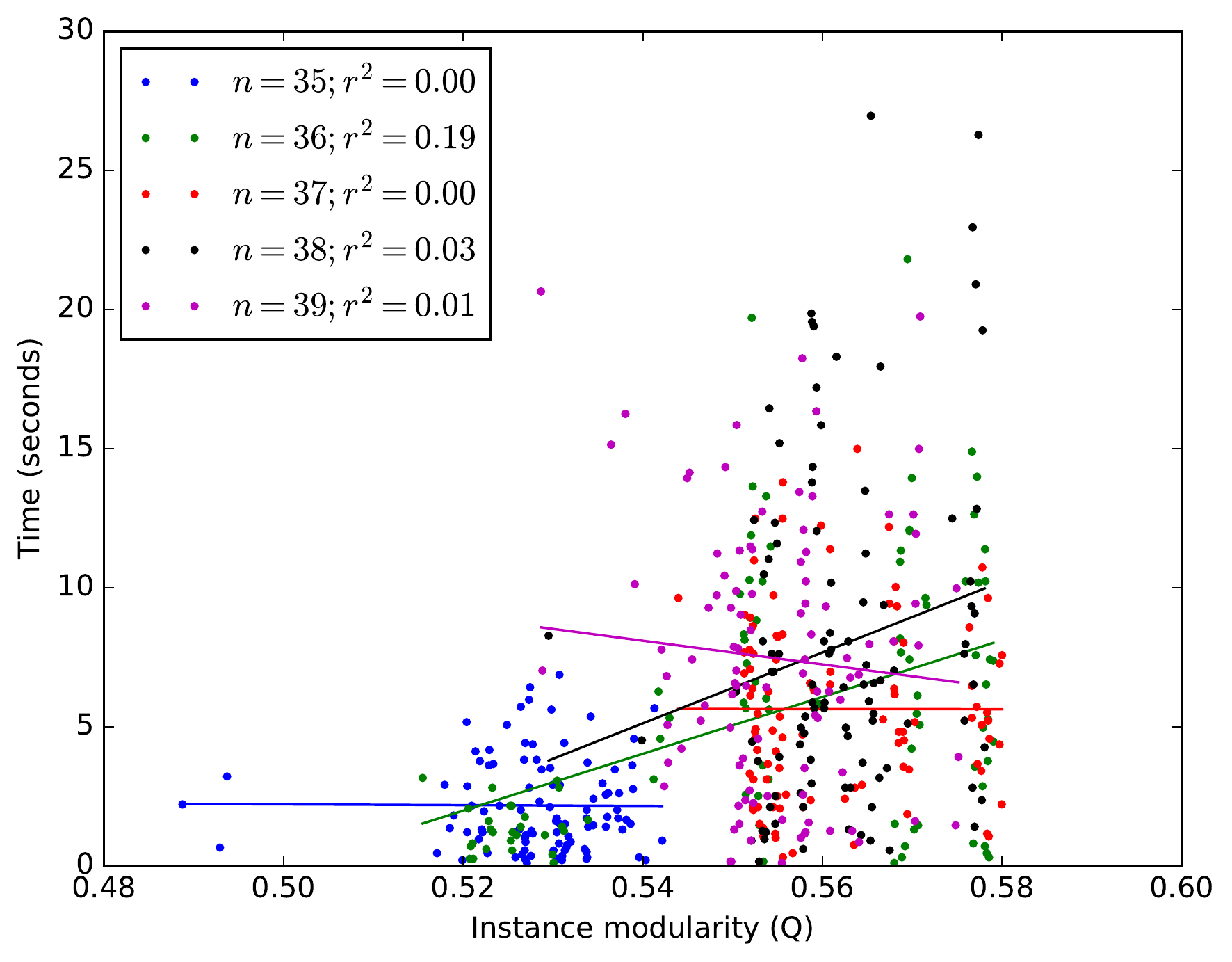}
  \caption{Solver times per community structure. We only display the results
    for some values of $n$ to avoid more clutter, but similar results
    hold for all $n$.}\label{fig:commstruct}
\end{figure}
\fi 

Even after this preprocessing step the instances for long
multiplication have too little variation to conclude anything about
the relation between $Q$ and solver time.  For some Karatsuba instances
the results are given in
\autoref{fig:commstruct}.  The results are grouped according to the
bitlength $n$ and per group linear regression is applied to each group.
The low $r^2$-values suggest there is no relation between modularity
and solver time for these instances.

Interestingly, the values of $Q$ are relatively high and far outside
the range $0.05 \leq Q \leq 0.12$ for which the instances were
conjectured to be hard, yet the instances are still hard to solve.
The above data leads to the conclusion that the community structure does
not provide a good prediction for solver runtime when applied to SAT
instances that encode multiplication circuits.

\subsubsection{Other metrics}

Besides the above metrics that can be computed on any SAT instance,
one might consider if there is any correlation between metrics that
apply only to this specific use case.  
In particular we are interested if there is any pattern in $N$, $p$
and/or $q$ that the solver is able to exploit for a faster solving time.
Since SAT instances are defined over Boolean variables we considered
the Hamming weight of: $N$, $p$, $q$, and $p \oplus q$.
We also measure if the solver is able to pick up on some patterns that
make a number easier to factor according to number theoretic methods
(such as Pollards $p-1$ method):
smoothness of $p-1$,
smoothness of $q-1$,
$|p-q|$, and
$\log N$.
We measured the correlation with the solver time (see
\autoref{app:patterns} for details).  No metric shows any significant
correlation.

\subsection{Comparison to number-theoretical methods}

One can put the above results in context by comparing the absolute
runtime to that of other number-theoretical results.
Using SageMath~\cite{sagemath} we measured the runtime of two
approaches: factoring with the built-in factor function:~\autoref{fig:sagefactor}
and factoring by trial division:~\autoref{fig:trialdiv}.

\ifacm
\begin{figure}
  \includegraphics[width=\columnwidth]{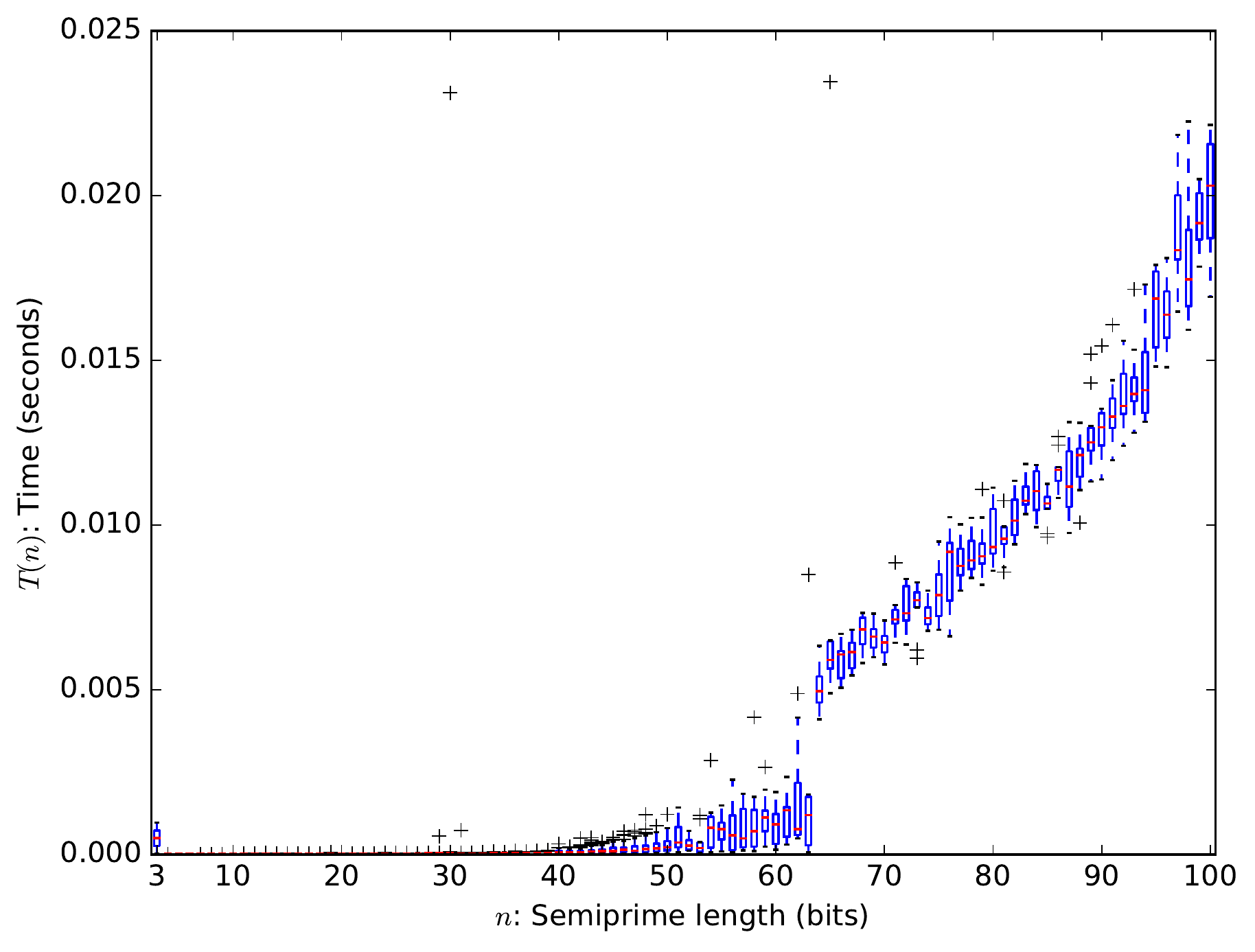}
  \caption{Runtime of SageMath \texttt{factor()}}
  \label{fig:sagefactor}
\end{figure}
\begin{figure}
  \includegraphics[width=\columnwidth]{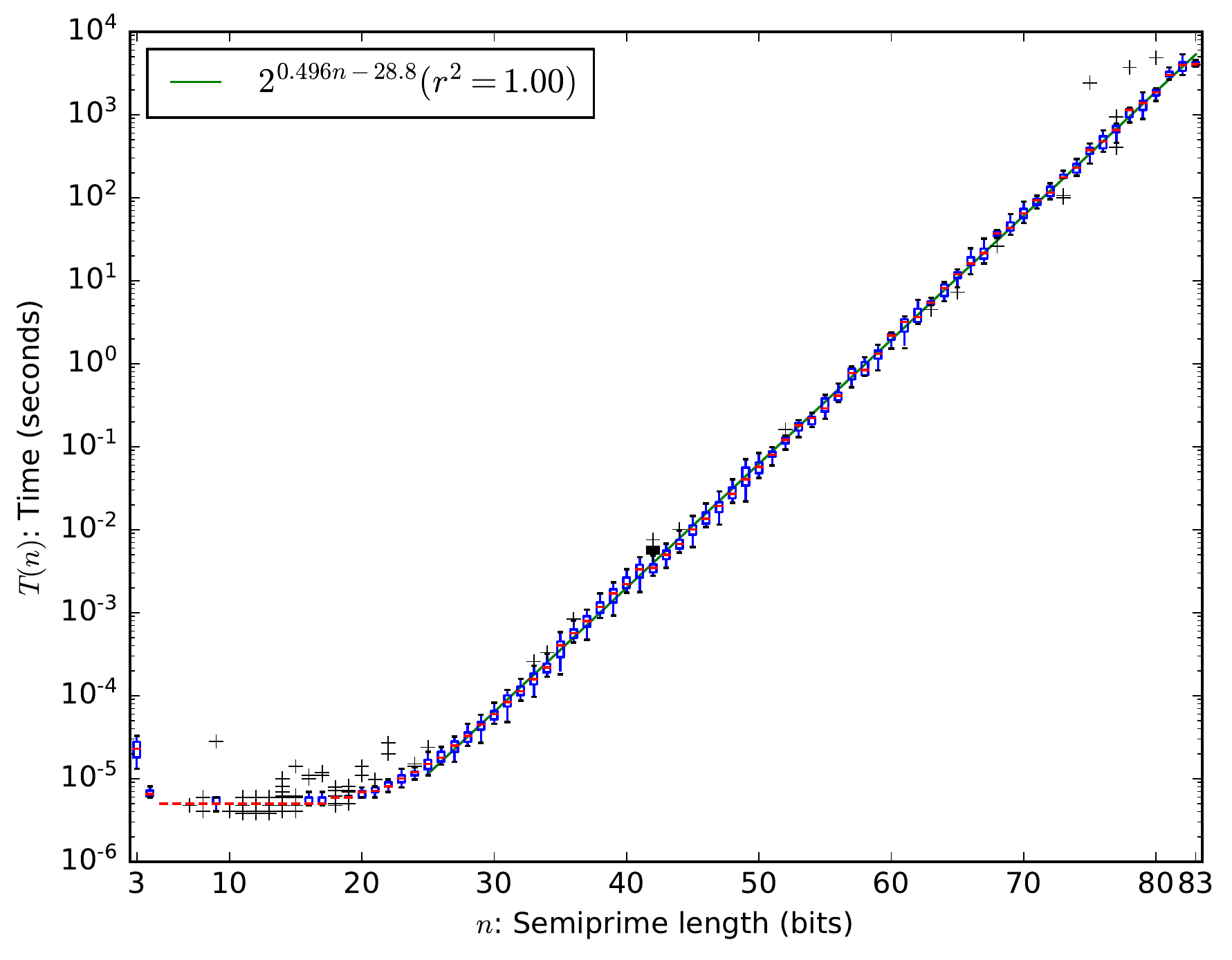}
  \caption{Runtime of SageMath trial division}
  \label{fig:trialdiv}
\end{figure}
\else 
\begin{figure}[ht!]
  \centering
  \begin{subfigure}[t]{0.49\textwidth}
    \includegraphics[width=\textwidth]{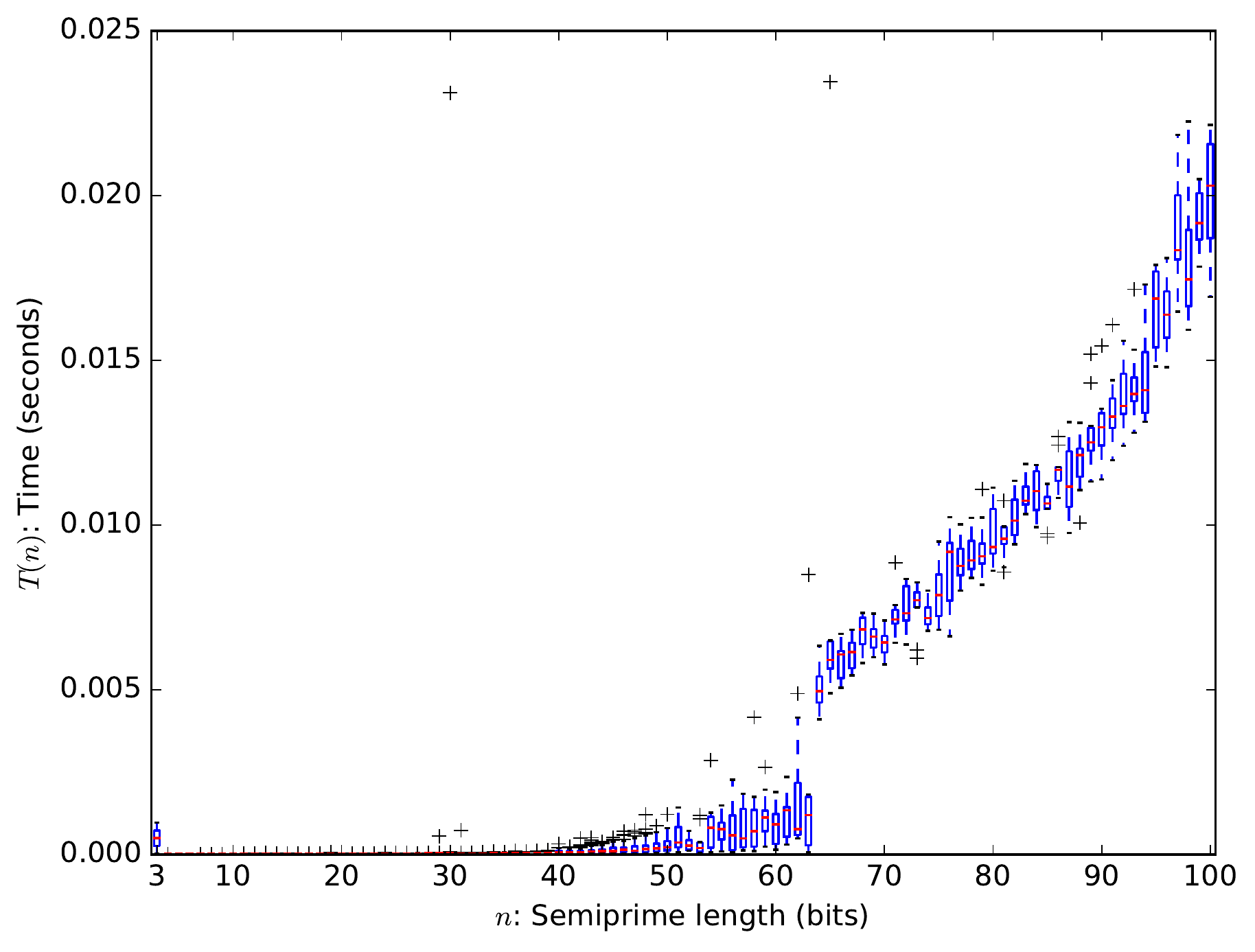}
    \caption{SageMath \texttt{factor()}}\label{fig:sagefactor}
  \end{subfigure}
  ~
  \begin{subfigure}[t]{0.49\textwidth}
    \includegraphics[width=\textwidth]{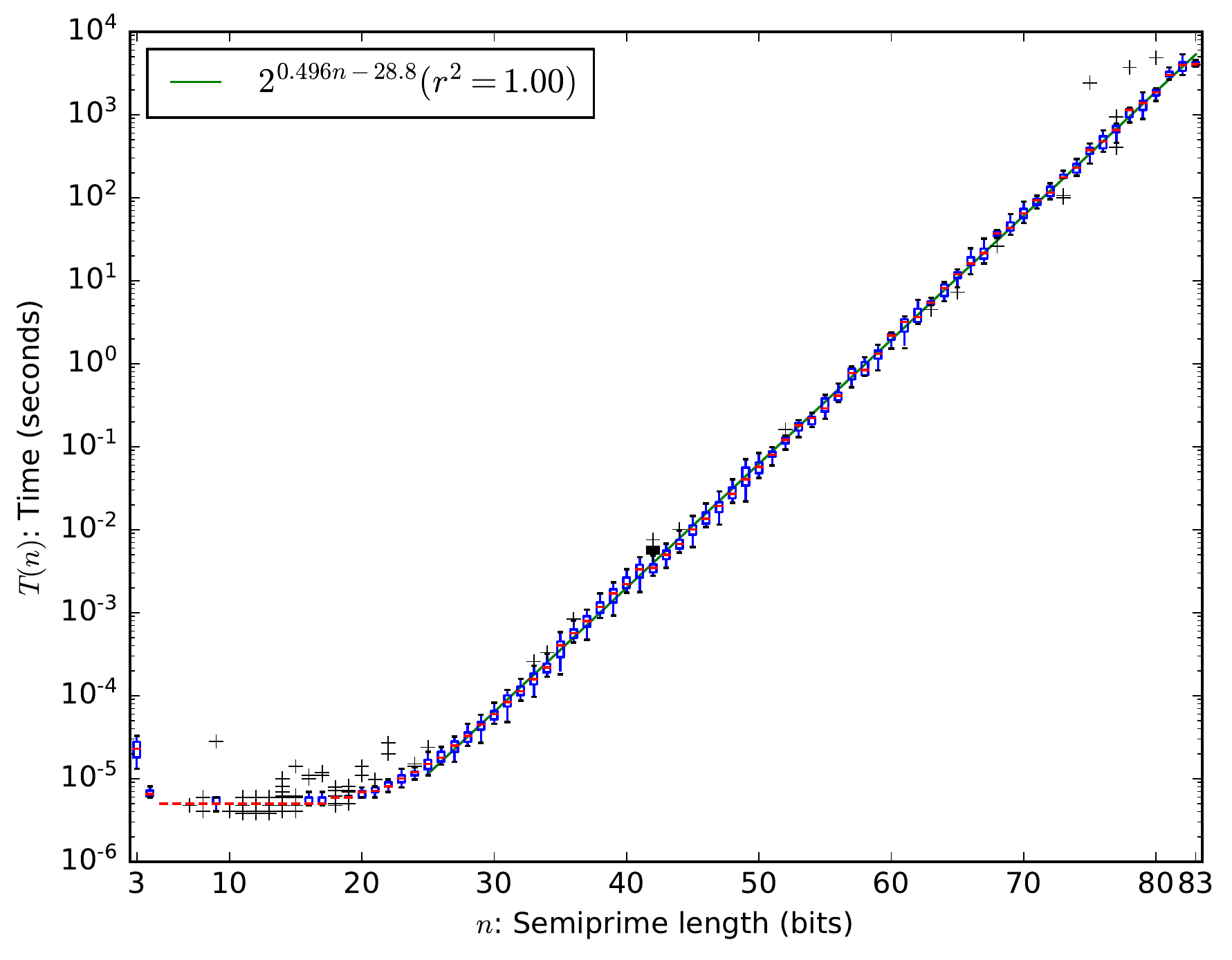}
      \caption{Trial division}
    \label{fig:trialdiv}
  \end{subfigure}
  \caption{Runtime of factoring using numerical methods. No randomization was applied for obtaining these results.}\label{fig:num}
\end{figure}
\fi 

SageMath is able to factor almost all semi-primes up to a 100 bits in
under 0.025 seconds.
The tested semi-primes are so small that
the asymptotic behavior of the underlying algorithm is not even
visible yet, so there is no point in extrapolating these results.
In fact the crossover point where the number field sieve (NFS)
is faster than asymptotically slower methods such as the quadratic
sieve and the elliptic-curve method is much larger than 100 bits,
so that SageMath is not even using NFS to factor these small numbers.
Instead, we refer to the literature to find that the best classical
factoring algorithm (the general number field sieve~\cite{llmp93})
runs in $L_N[1/3, {(64/9)}^{1/3}]$ and this was indeed used to factor a 768-bit
RSA modulus in approximately two-thousand core-years~\cite{rsa768}, with one core being
a 2.2~GHz AMD Opteron.

The timing of factoring using trial division is shown in
\autoref{fig:trialdiv}.  The results reveal an exponential trend and
with a much smaller constant than the SAT solver.  On this
small scale on which measurements were performed, trial division
easily outperforms the SAT solvers.  The asymptotic runtime of the 
methods are so
close together that we cannot meaningfully extrapolate the results
to find a cross-over point where the SAT solvers become faster than
trial division.  We therefore cannot rule out that factoring with
classical SAT solvers is always slower than trial division.

\section{Quantum Solvers}\label{sec:quantum}

State of the art classical factoring algorithms have
super-polynomial runtime $L_N[1/3, (64/9)^{1/3}]$~\cite{llmp93}, whereas Shor's
algorithm~\cite{shor94} runs in polynomial time.  This algorithm
requires a fault-tolerant quantum computer and no scalable version
has been implemented yet.  Shor's algorithm has
profound practical implications for currently deployed public-key
cryptography such as RSA and the timing of the factoring of 1024-bit,
2048-bit or even larger semi-primes is of great practical significance
for both contemporary and future security systems~\cite{mosca13}.
Mitigations for future systems and current systems requiring long-term
security are being researched by the field of post-quantum
cryptography~\cite{pqc09, nist, nistpqc}.

An interesting notion of quantum computing has been proposed
by Farhi et al.~\cite{farhi00} in the form of adiabatic quantum computers.  
It was suggested that adiabatic quantum algorithms may be able to outperform
classical computers on hard instances of NP-complete problems~\cite{fggllp01}.
Since then,
adiabatic quantum computation (a generalization of the adiabatic optimization
explored deeply by Farhi et al.) has been proven to be polynomially
equivalent to quantum computation in the standard gate
model~\cite{adkllr04}.
While the possibility of super-polynomial (or even just super-quadratic)
quantum speed-up for NP-hard problems remains an open question\footnote{it
is known that any such speed-up must go beyond pure ``black-box''
search~\cite{bbbv96} as attempted by Farhi et al.~\cite{fggllp01}
and must somehow exploit additional knowledge or structure~\cite{dmv01}}
it is generally believed that quantum
computers (including adiabatic quantum computers) are not able to
efficiently solve NP-hard problems such as SAT.
Note that this assumption is implicit, e.g. in the fact that post-quantum
cryptographers are working on the assumption that symmetric algorithms like AES
and SHA that offers $n$ bits of security against classical attacks offer $n/2$
bits of security against the best known quantum attacks~\cite{nist}.\footnote{
  Excluding some specific attacks in the ``quantum superposition'' attack
model~\cite{superpos-attack}.}
  In this section we consider the speedup that can be
achieved by reducing the problem of factoring a semi-prime to an instance
of an NP-hard problem which is then solved with a quantum computer.

When considering the runtime $T$ of an algorithm we are most
interested in the runtime as a function of the input size.  In order
to determine if one solver is faster than the other, we should always
consider the \emph{total} runtime.  In the above analysis this is what
we did by measuring the total runtime of the SAT solver
\emph{including} the runtime of the preprocessor.\footnote{To be even
  more precise we should also have included the time it took to
  generate the SAT instances.  This generation is done in polynomial
  time and the runtime is negligible compared to the solver time,
  therefore we omitted this from our benchmarks.}  For
many solvers the total runtime can be naturally partitioned into the
time spent in pre-/post-processing ($T_p$) and the time spent solving
($T_s$):
\begin{equation}
  T(n) = T_p(n) + T_s(n),
\end{equation}
where $n$ is the input size of the problem.

Examples of this partitioning occur with the SAT preprocessor ($T_p$) and the SAT solver ($T_s$),
the compiling ($T_p$) and running ($T_s$) of Shor's algorithm or the
creation of a Hamiltonian ($T_p$) and the execution of the adiabatic
algorithm ($T_s$).

In order to properly analyze the runtime of any algorithm we need
to consider $T(n)$ and not just $T_s(n)$, since an unbounded amount of
preprocessing can find a solution and render $T_s(n)$ to be trivial.
We should also take care to set $n$ to be the input size
of the problem.  Concretely this means we should let $n$ be the size of the
semi-prime and not the number of variables or clauses in our SAT instance.
It is also important to analyse instance sizes larger than
some lower bound ($n \geq n_0$), as the asymptotic behaviour is 
not visible for smaller sizes.  For example the asymptotics of the MapleCOMSPS
solver on integer factorization only become apparent at $n_0 = 20$ bit
semi-primes.

\subsection{Faster SAT solvers}\label{ssec:speedup}

One might hope that we can apply a quantum strategy that can improve
on the best known classical methods.  We chose SAT solvers to
represent the best classical methods as their implementations are the
highly optimized result of years of research.
Generic quantum searching methods can achieve at most a quadratic speed-up, and
we are aware of no convincing evidence that more than a quadratic speed-up can
be achieved by quantum SAT-solving methods.
For example,
many modern SAT solvers rely on machine-learning
techniques~\cite{maplecomsps} and many quantum methods with a
quadratic speedup are known for a variety of machine-learning
algorithms~\cite{bwpr17}.
See also~\cite{aar15} for why the
exponential speedup promised in some quantum machine-learning literature is
unlikely to be achieved in real-world implementations.

A quick calculation shows that even with a quadratic speedup, this
strategy is not a very efficient one.  We set an upper bound on the
number of operations required for the classical solver based on our
results.  Accounting
for any internal parallelism in the processor (four arithmetic ports
per processor) and assuming that the CPU was fully occupied at every
clock cycle this means that $10^{10}$ operations were being executed
every second during the solving time.

\ifacm
\begin{figure}
  \includegraphics[width=\columnwidth]{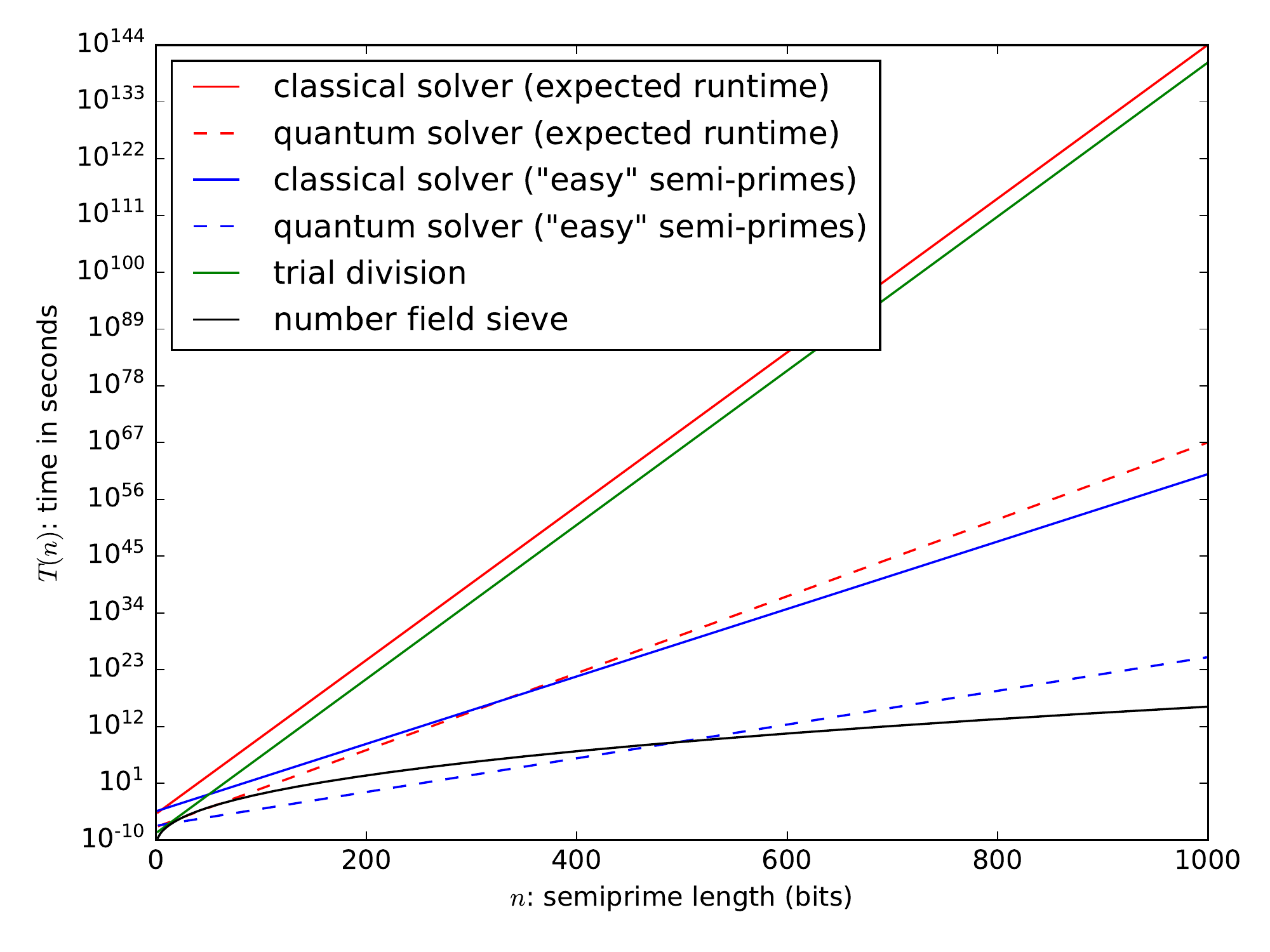}
  \caption{Comparison of efficiency of various factoring methods. The classical results
  are extrapolated from experimental data. The quantum results apply a quadratic speedup
  over the full classical computation. The number field sieve result plots $L_N[1/3, (64/9)^{1/3}]$
  operations assuming the same number of operations per second.
  }
  \label{fig:combined}
\end{figure}
\else 
\begin{figure}[ht!]
  \centering
  \includegraphics[width=.6\textwidth]{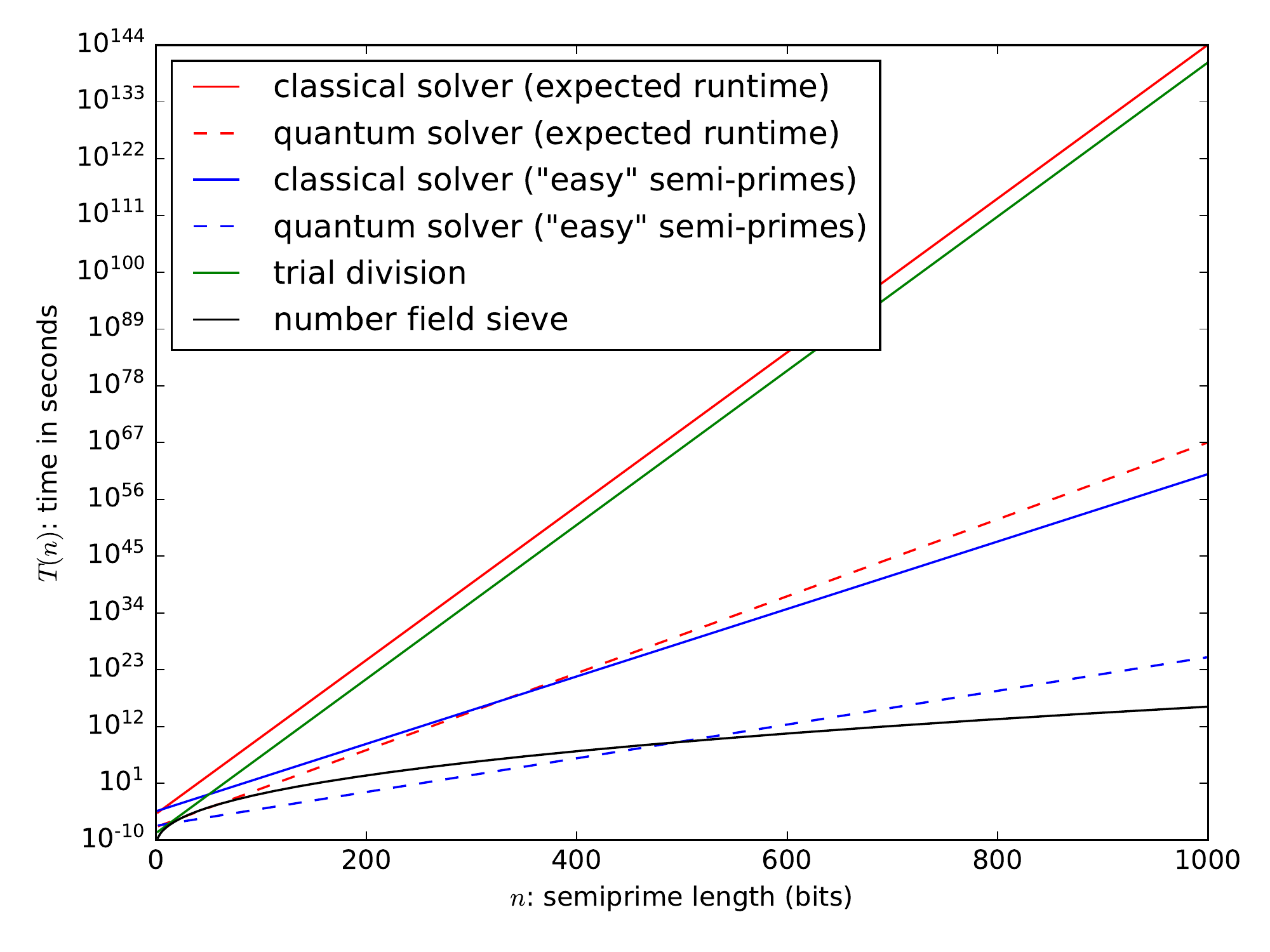}
  \caption{Comparison of efficiency of various factoring methods. The classical results
  are extrapolated from experimental data. The quantum results apply a quadratic speedup
  over the full classical computation. The number field sieve result plots $L_N[1/3, (64/9)^{1/3}]$
  operations assuming the same number of operations per second.
  }\label{fig:combined}
\end{figure}
\fi 

Under this assumption the expected number of operations required for
classical SAT solving becomes ${2^{16.8}\cdot2^{0.495n}}$.  With a
quantum computer we might hope to reduce this to
$\sqrt{2^{16.8}\cdot2^{0.495n}} = {2^{8.41}\cdot2^{0.247n}}$ operations.
To put
this in perspective, consider a quantum computer that can
execute $10^{40}$ \emph{quantum} operations per second.
Note that even a classical computer with such speeds could break AES-128, 
SHA-256,
RSA-2048 and
ECC P-256 in an instant, so this is a very generous upper bound.
Under these assumptions it
would still take approximately a hundred times the lifetime of the
universe to factor the RSA-768 number using the quantum SAT solving approach,
whereas this number has been factored classically on a real machine
in two-thousand core-years using number-theoretical methods.
A visual comparison of these results are given in \autoref{fig:combined}.
 
Note that all estimates so far are biased towards more (classical)
operations per second.  The reason is that we want to compute an upper bound on
the speedup that can be achieved by applying Grover's algorithm (or
some alternative quadratic speedup) in
order to factor numbers with SAT solving.  It is almost certain that the
processor executed less operations and it is very unlikely that
the quadratic speedup can be applied to the full computation without any
overhead of executing the algorithm.  Therefore, classical algorithms
will likely require less operations than reported and quantum
algorithms will likely require more operations than computed.

Note also that the known speedups for quantum solvers
are applied to $T_s$, even though the above calculation
generously assumes that $T_p(n) = 0$ and the
speedup can be applied to the full calculation time $T(n)$.  For most
classical solvers it indeed holds that $T_p(n) \ll T_s(n)$ as $n$
grows large enough, but for many of the adiabatic factoring methods discussed
below it holds that
$T_p(n) \gg T_s(n)$ as $n$ grows.  This means that our calculation is
a significant overestimation of the maximum speedup that can be achieved
with the adiabatic factoring method.

\subsection{Adiabatic factoring}

The method used for factoring with the adiabatic algorithm first
reduces factorization to finding the roots in a set of integer
equations in which the unknown variables are restricted to binary
values, corresponding to the input bits of the prime and carry bits of
the intermediate computation.  This is translated to the
pseudo-Boolean optimization problem by squaring all equations (so that
the roots correspond to the minimum values) and summing over all
equations.
This reduction was first suggested by Burges as a method for
generating unconstrained optimization problems whose complexity can be
easily controlled~\cite{bur02}.
The adiabatic algorithm~\cite{farhi00} is particularly well
suited for encoding optimization problems of this kind: the resulting sum
describes a Hamiltonian of which the ground state encodes the solution
and every variable corresponds to a single qubit.  In general it is
not easy to physically initialize the system in the ground state of
the Hamiltonian, so instead an easier Hamiltonian encodes the initial
state which is easy to initialize in the ground state.  The adiabatic
theorem tells us that if we evolve the physical system from the
initial Hamiltonian to the final Hamiltonian slow enough, the system
will remain in its ground state.  Measuring the final state will then
provide the answer to the optimization problem.

To assess the power of the adiabatic algorithm it is therefore
important to quantify how fast this evolution can be done.  A coarse
lower bound is given by the spectral width of the time-dependent
Hamiltonian, but sharper bounds on the runtime so far elude
us~\cite{dmv01}.
This has led some researchers to study the applicability of the adiabatic
algorithm to some NP-complete problems~\cite{fggllp01}.
Most evidence for a speed-up is
based on noise-free simulations on small instances (for which the
asymptotic behaviour might not be visible) which are chosen randomly,
shedding light on typical performance for small instances.
Cryptographic problems require
average-case hardness in order to be practical, which is why they are
so suitable for testing worst-case behaviour of algorithms that solve
them, especially when the reduction to an NP-hard problem is as simple
as reducing factoring to SAT as demonstrated in the previous section.

Pseudo-Boolean optimization is known to be NP-hard, meaning amongst
other things that a polynomial reduction exists from 
the SAT problem.
The objective function for factorization instances using the above
method is a quartic polynomial.  Real-world demonstrations of the
adiabatic algorithm suffer from additional limitations (besides
noise-resistance) in the number of available qubits and multi-qubit
interactions.  The latter limitation means that quartic terms in the
objective function cannot always be realized.  Using
quadratization~\cite{rosenberg75} each objective function can be
simplified to a quadratic polynomial at the price of additional
variables, giving an instance of the well-studied quadratic
unconstrained binary optimization (QUBO)\footnote{also known as
  unconstrained binary quadratic programming (UBQP)} problem.  This
simplification runs in polynomial time and results in only
polynomially many variables overhead, so the problems are equivalent.

However for many real-world systems the extra variables (qubits) are
not available, so additional simplifications are required.
This is fine as long as these simplification steps do not dominate
the overall runtime of the program.  More precisely we can execute polynomially many
simplification operations and $T_p(n)$ will remain polynomial in $n$, thereby
not significantly increasing the runtime $T(n)$ which is dominated
by the super-polynomial runtime $T_s(n)$.
When the simplification process is allowed to have an exponential runtime
it can absorb the hardness of the problem, leaving a weaker
problem to be solved (trivially) in polynomial time.

\subsubsection{Implementations}

The first adiabatic factorization~\cite{peng08} was implemented in
2008 using nuclear magnetic resonance (NMR) to factor 21 using three qubits.  The authors
fit a quadratic curve against a theoretical approximation in a
noiseless model, they measure the runtime as a function of the
number of input qubits (not the size of the factored number) and
they only consider the small domain of seven to sixteen input qubits.
It is doubtful that such small instances are a good indicator of polynomial asymptotic behaviour.

Later work~\cite{xzlzpd11} translates the problem of factoring 143
into a pseudo-binary optimization instance, which is an NP-hard
problem~\cite{bh02}.
The
authors manage this by introducing the additional assumption that both
factors must be of equal bitlength with the most significant bit set to
one.  Combining these assumption with some simplifications in the
pseudo-Boolean equations simplifies the problem so that it only
concerns four input bits of the prime factors.  Although the
used simplifications are efficient, only an upper bound of their effectiveness 
is given.

Subsequent research~\cite{db14} observes that a minor generalization
of the previous method reduces the problem to four input qubits
whenever the two primes composing the semi-prime differ only in two
positions, which likely occurs for infinitely many semi-primes~\cite{polymath14}.
This provides some evidence that the simplifications
do not generalize and the factored number 143 was identified as
a particularly easy number to factor.
In other words, this example was hand-picked from an exponentially unlikely
family of semi-primes that are by design easy to factor.
The authors report the number 56153 as being the largest semi-prime factored
quantumly and at the same time argue that the work has factored an arbitrarily
large set of semi-primes (since they can be pre-processed into solving the same
pseudo-Boolean equations).
The
reason for not reporting a bigger number appears to be the large runtime $T_p$ of
the simplification process.

Much subsequent research in the adiabatic factoring field has focussed
on methods such as deduc-reduc~\cite{tod15}, split-reduc~\cite{otd15}
and energy landscape manipulation~\cite{tld15}, all of which can be
seen as improvements on the preprocessor runtime $T_p$ and none of
which do any improvements on $T_s$.

The problems with viewing these works as relevant quantum integer
factorization benchmarks is highlighted even further in the more recent
paper that claims to have factored 291311 with adiabatic quantum
computation~\cite{li17}.  The authors take the above approach and
reduce the problem of factoring 291311 to the integer equations
\begin{align}
  q_1 + q_2 - 2 q_1 q_2 &= 1 \\
  q_2 + q_5 - 2 q_2 q_5 &= 0 \\
  q_1 + q_5 - 2 q_1 q_5 &= 1,
\end{align}
where the variables $q_i$ must take on binary values and represent
unknown bits in the binary representation of factor $q = 1000 q_5 01
q_2 q_1 1$.  The authors stop their simplification process at this
point and fail to notice that the above equations can be further
simplified to
\begin{align}
q_1 = 1 - q_2 = 1 - q_5.
\end{align}
Both solutions $q_1 = 0$ and $q_1 = 1$ correspond with respective
factors $q = 557$ and $q = 523$.  In other words, the number was
already factored by the simplification process and the adiabatic
quantum computation was a complicated way of flipping
a coin and deciding between the two factors.
The above criticism of these claims to meaningful quantum factoring benchmarks
was in fact already made in 2013~\cite{ssv13b}.

A method called Variational Quantum Factoring (VQF)~\cite{aoac18}
employs the same strategy for factoring, which is to reduce it to an NP-hard optimization problem.
The authors are careful to ensure that preprocessing takes only polynomial time.
Although the authors claim that ``VQF could be competitive with Shor's algorithm even
in the regime of fault-tolerant quantum computation'', we find no convincing
argument to support this conjecture.
In particular,
they do not provide convincing evidence that the solving step is efficient:
no semi-prime larger than $2^{15}$ is considered by their work and they observe that
``the mere presence of carry bits negatively affects the algorithm''.

The criticism from~\cite{ssv13b} applies equally well
against ``compiled versions'' of Shor's
algorithm: both implementations require much
precomputation and therefore do not scale to factoring larger numbers.
The problem is that both precomputations require prior knowledge of
the solution.
``Compiled versions'' of Shor's algorithm were never intended 
to scale to meaningful input sizes, as is highlighted in
the abstract of the work factoring 15 with NMR:
``scalability is not implied by the present work.  The significance
of our work lies in the demonstration of experimental and theoretical
techniques''~\cite{vsbysc01}.

The important difference is that the runtime of Shor's algorithm is 
well understood and provides a super-polynomial speedup in $T_s$ over even the
best numerical methods for factoring.
As fault-tolerant hardware emerges, we can simply strip away the non-scalable optimizations.
On the other hand the runtime
of reducing factoring to an NP-hard problem and then solving
it with (quantum) solvers is not understood very well, but all
evidence points in the direction that it cannot even compete with 
classical
numerical methods for factoring.

\subsection{D-Wave}\label{ssec:dwave}

The D-Wave systems work by a process called quantum annealing, which
can be viewed as a noisy version of adiabatic quantum computing.
It has been shown that $O(n^2)$ qubits suffice to encode factoring into a
quantum annealing instance with local interactions~\cite{dwavepat}.
The article ``Boosting integer factoring performance via
quantum annealing offsets''~\cite{dwave-int}
describes a ``boost'' when comparing factoring on the D-Wave machine with
annealing offsets against the D-Wave machine without annealing offsets.
The largest factored number has 20 bits.

All semi-primes up to 200000 (18 bits) have been factored with help of the D-Wave 2X
by heuristically mapping the optimization problem to the Chimera graph
underlying the machine~\cite{da17}.
Exponential methods from computational algebraic geometry are used for
preprocessing the instances without quantification of the (asymptotic
or measured) runtime so that there is no indication of the efficiency
of this preprocessing step.
Although some statistics on the annealing process are provided for six semi-primes,
not enough information is given for a meaningful assessment on the scalability
of both the efficiency and effectiveness of this method.

Integer factorization has been implemented on the D-Wave 2000Q by
a similar strategy~\cite{jbmhk18}.
Quantified experimental results are only provided for factoring
15 and 21.  As the authors note, there
is no evidence that quantum annealing will find factors with significant
likelihood in polynomial (or even sub-exponential) time.

\section{Conclusions}\label{sec:conclusions}

SAT solvers are not known or believed to be able to factor semi-primes efficiently.  Overall,
even the fastest solver (MapleCOMSPS) has an exponential runtime in
the size of the factors.  Closer inspection of the solver runtime
indicates that the solver is not able to detect any pattern in the SAT
formulas that encode the factorization problem.
Asymptotically the solver runtime appears to be 
comparable to that of trial division, but this advantage is almost completely
negated by the overhead in the constant term.  The performance of SAT
solvers does not even come close to that of number-theoretical
methods.

Quantum SAT solvers are not expected to do much better.  Even when
calculating a very optimistic speedup to the current state-of-the-art
classical solvers, these solvers are outperformed with orders of
magnitude by (classical) number-theoretical factoring methods.  This
approach to factoring reduces factoring, a problem with an $L_N[1/3, (64/9)^{1/3}]$
algorithm, to an NP-hard problem and then running (classical or
quantum) solvers that have exponential runtime in the worst-case. At the surface, this
obviously does not sound like a 
promising idea, as the quantum SAT solver must make up the exponential
ground lost by translating the problem with subexponential algorithms
to one where the best known algorithms are exponential.
One might hope that good SAT solving heuristics for solving SAT on
random or average-case instances could nevertheless have a practical impact
on integer factorization, but there is no evidence of this.
Of course if it were that easy RSA would be broken
regularly by SAT solvers which does not appear to be the case.
Furthermore, in practice it appears that SAT instances derived from integer factorization instances are hard SAT instances.
Thus it would be especially surprising if a SAT solver of any kind
(quantum or classical) could solve these instances with resources
comparable to that of using the classical number field sieve (i.e.
subexponential complexity).
Our
work explores this possibility more deeply and reinforces the folklore
that reducing multiplication to SAT and then applying SAT solvers,
classical or quantum,
is not useful for factoring numbers of sizes relevant to cryptography.


A more promising approach is to try to speed up the solution to some subroutine
of the NFS, as is done in~\cite{bbm17}. In particular, one could reduce some
carefully chosen sub-problem solved within the number field sieve to SAT.
The sub-problem should be chosen so that classically solving the SAT
instance is roughly as costly as the usual approach to solving the sub-problem.
In this case, any quantum speed-up for solving these SAT instances would lead
to a faster implementation of the number field sieve.
This approach is explored in~\cite{mvv19}.


Of course, one cannot rule out unexpected breakthroughs in quantum SAT solving
or a wide range of other quantum or classical approaches to factoring
semi-primes. However, it is important to distinguish the possibility of
unexpected breakthroughs (especially those that contradict conventional wisdom
or lack a plausible roadmap) from tracking progress of an existing hardware
platform and of an algorithm that is pertinent for cryptographically relevant
semi-primes (i.e. classical computers and the NFS).

Once scalable
fault-tolerant quantum computers capable of implementing Shor's algorithm are
available, a similar tracking would be very meaningful (with the caveat
outlined in the introduction). In the meantime, it is important to track
progress toward achieving scalable fault-tolerant quantum computers.

In other words, notwithstanding other scientific merits of these works,
we are not aware of any evidence that any SAT-based quantum factoring results to date,
including factorization by quantum annealing,
are relevant milestones toward large-scale integer factorization
or the demonstration of a speed-up over the best known classical algorithms
for integer factorization.

\ifacm
\begin{acks}
We would like to thank Vijay Ganesh and Curtis Bright for the many
lessons about modern SAT solving and insightful discussions regarding
this project.
\end{acks}
\else 
\section*{Acknowledgment}
We would like to thank Vijay Ganesh and Curtis Bright for the many
lessons about modern SAT solving and insightful discussions regarding
this project.
We also would like to thank Colin P. Williams and
Kenneth Paterson for their helpful comments.
\fi 

\ifacm
\bibliography{paper}
\else
\sloppy
\bibliographystyle{plainurl}
\bibliography{paper}
\fi

\ifacm
\relax
\else
\clearpage
\fi

\appendix

\section{CryptoMiniSat 5}\label{app:cms5}

\ifacm
\begin{figure}
  \includegraphics[width=\columnwidth]{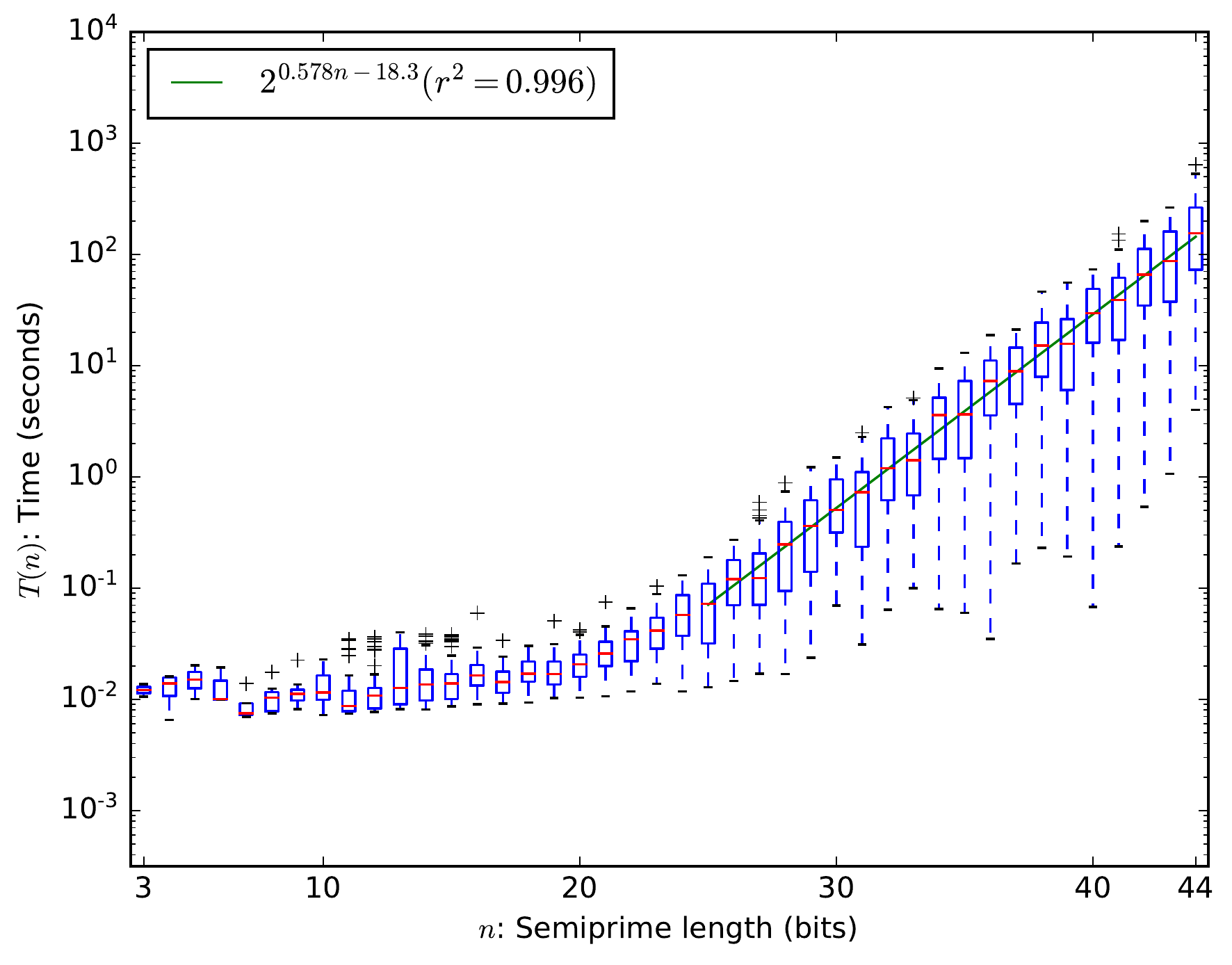}
  \caption{Runtime of CryptoMiniSat 5 on factoring semi-primes (schoolbook multiplication).}
  \label{fig:cryptolong}
\end{figure}
\begin{figure}
  \includegraphics[width=\columnwidth]{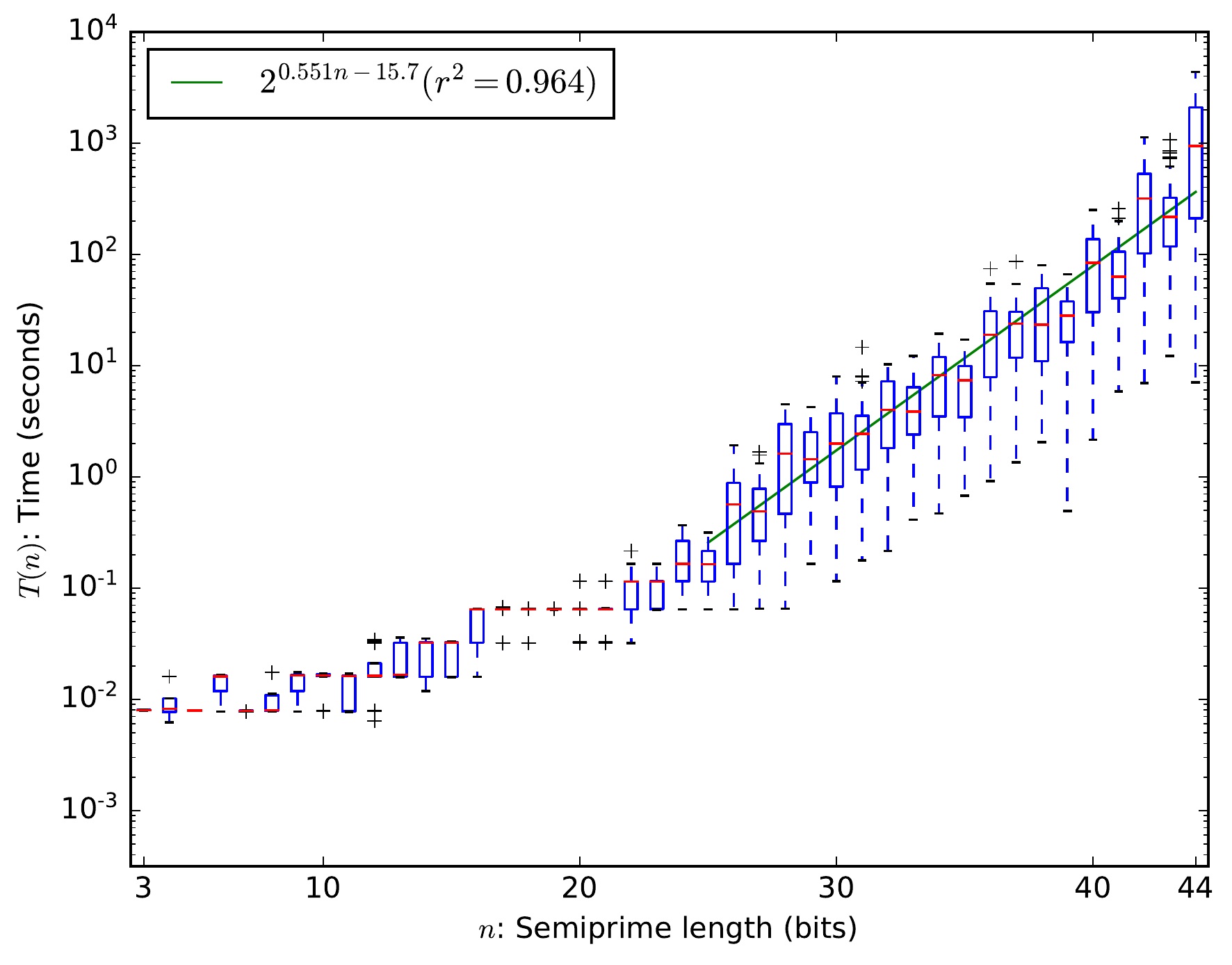}
  \caption{Runtime of CryptoMiniSat 5 on factoring semi-primes (Karatsuba multiplication).}
  \label{fig:cryptokar}
\end{figure}
\else 
\begin{figure}[ht!]
  \centering
  \begin{subfigure}[t]{.49\textwidth}
  \includegraphics[width=\textwidth]{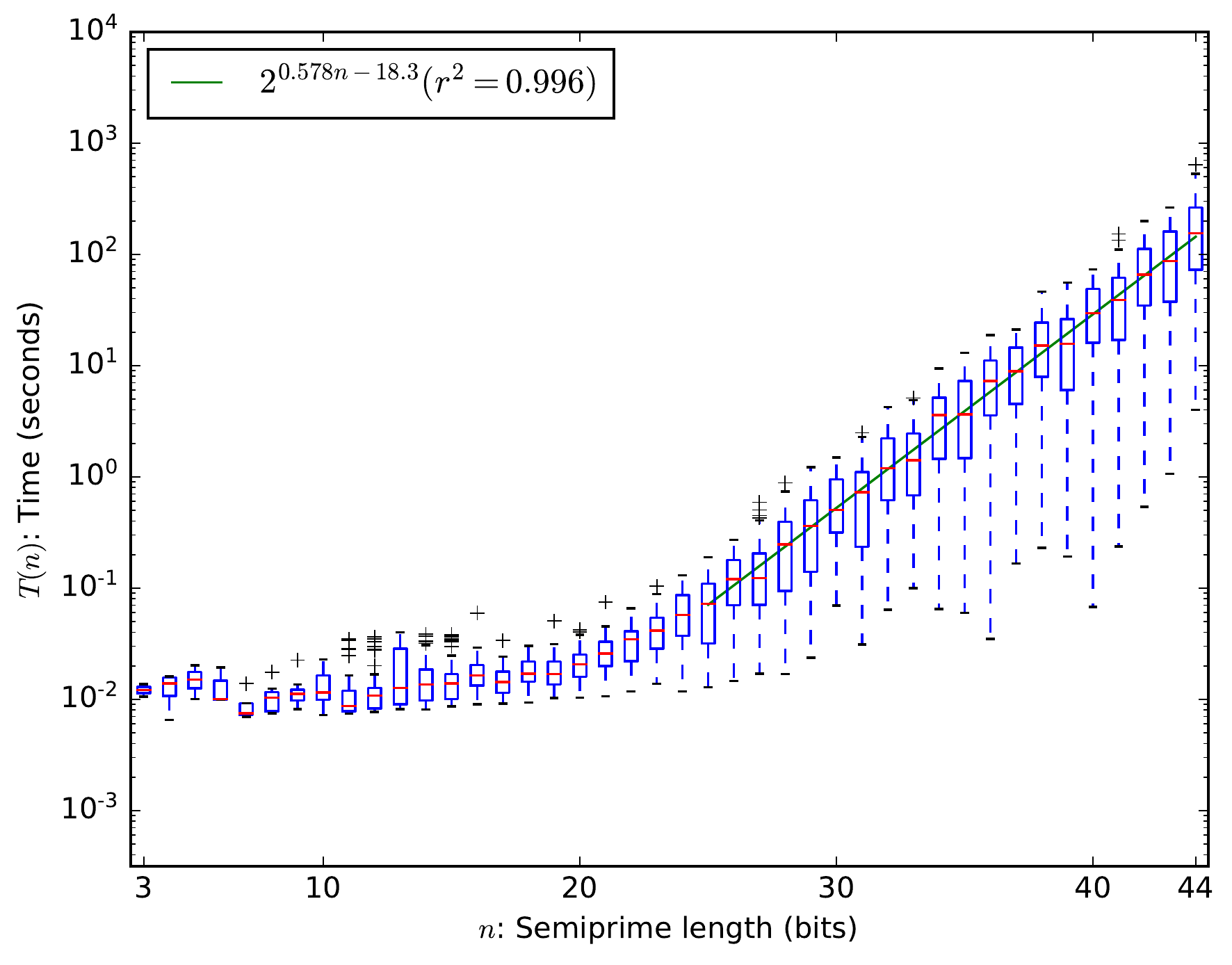}
  \caption{schoolbook multiplication}
  \label{fig:cryptolong}
  \end{subfigure}
  ~
  \begin{subfigure}[t]{.49\textwidth}
  \includegraphics[width=\textwidth]{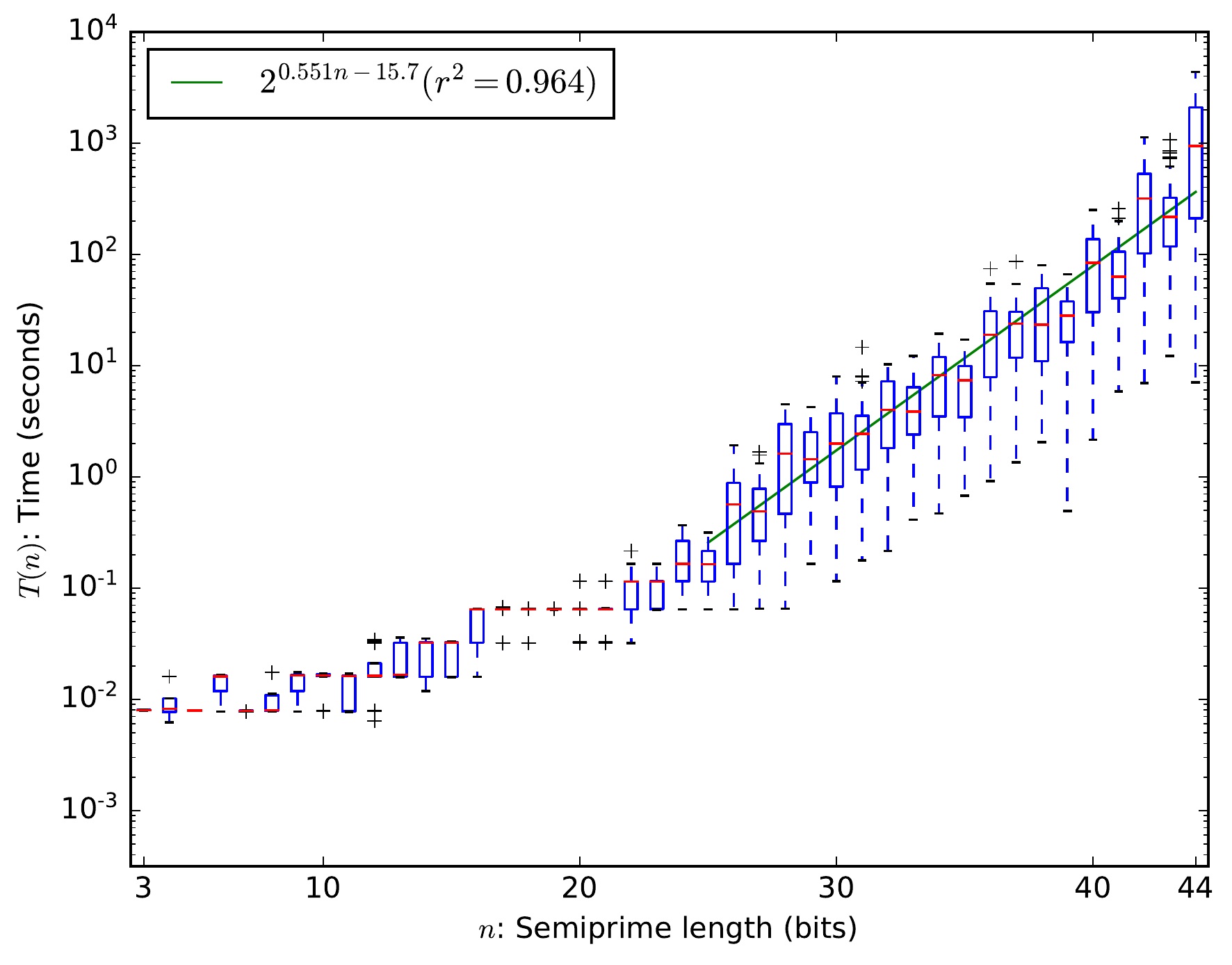}
  \caption{Karatsuba multiplication}
  \label{fig:cryptokar}
  \end{subfigure}
  \caption{Runtime of CryptoMiniSat 5 on factoring semi-primes.
    We measured 100 semi-primes per bitlength
    and applied no randomization.}
  \label{fig:crypto}
\end{figure}
\fi 

\autoref{fig:cryptolong} and \autoref{fig:cryptokar} show the performance
of factoring semi-primes with CryptoMiniSat 5.
This solver solved each semi-prime SAT instance
once, so no averaging has been applied to the shown results.
In particular, one might be tempted to
conclude from the longer whiskers in the depicted results that the
CryptoMiniSat solver is lucky more often.  However, the MapleCOMSPS
solver gives similar results when only considering one solution.
Closer inspection of the data reveals that CryptoMiniSat 5 is is
outperformed consistently by MapleCOMSPS.\@

\section{Patterns}\label{app:patterns}

\ifacm
\begin{figure}
  \includegraphics[width=\columnwidth]{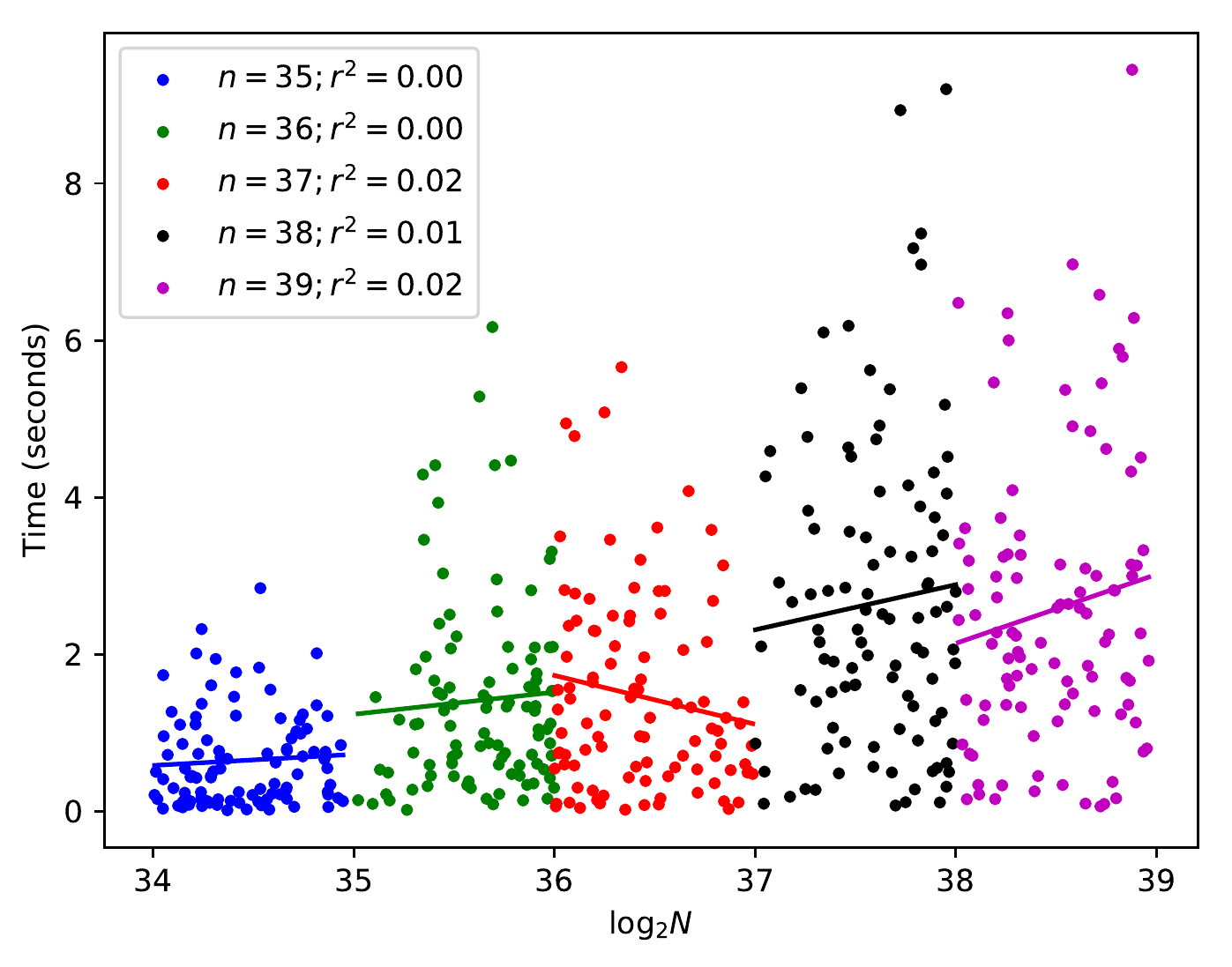}
  \caption{}\label{sfig:patlongsize}
\end{figure}
\begin{figure}
  \includegraphics[width=\columnwidth]{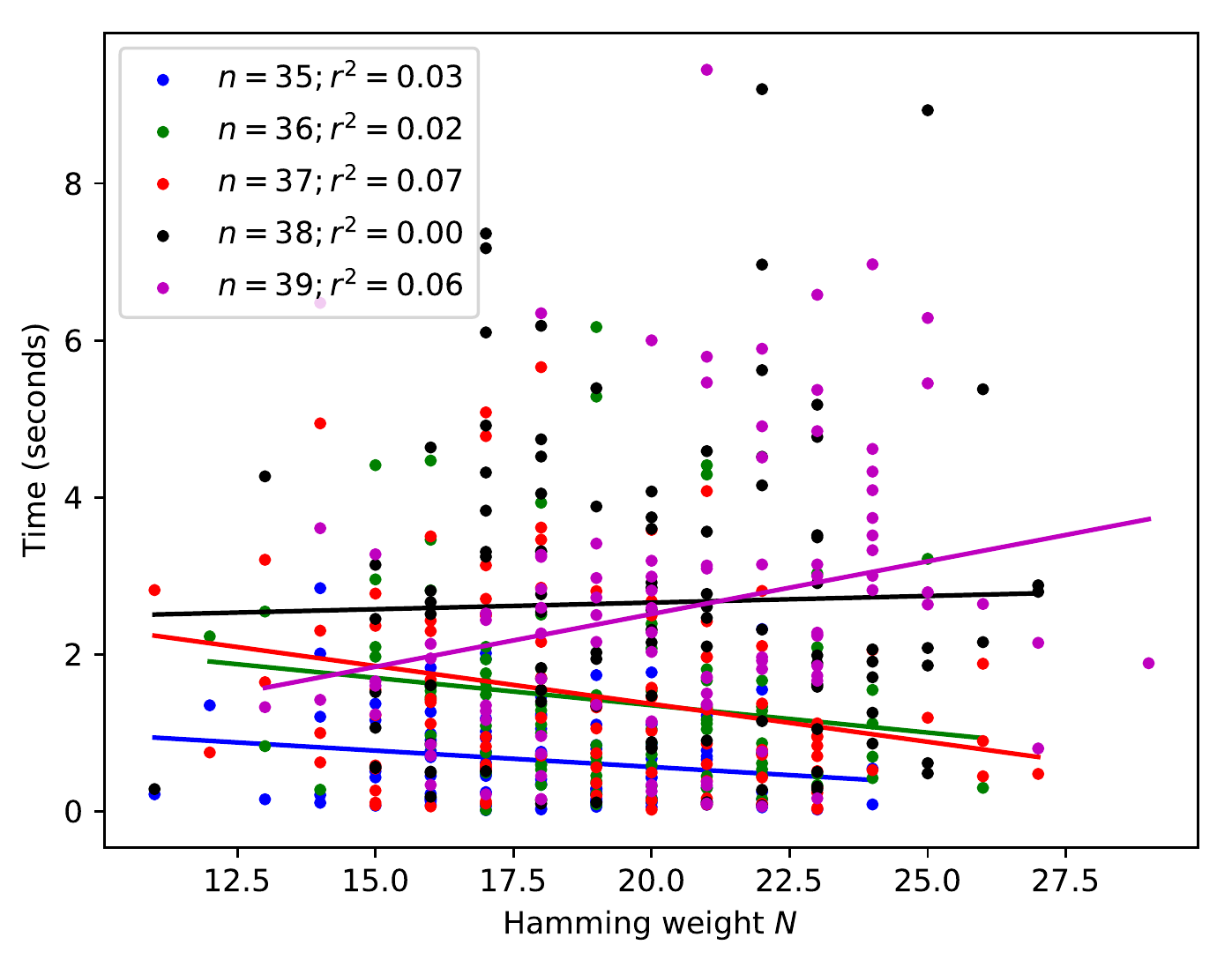}
  \caption{}\label{sfig:patlonghwn}                          
\end{figure}
\begin{figure}
  \includegraphics[width=\columnwidth]{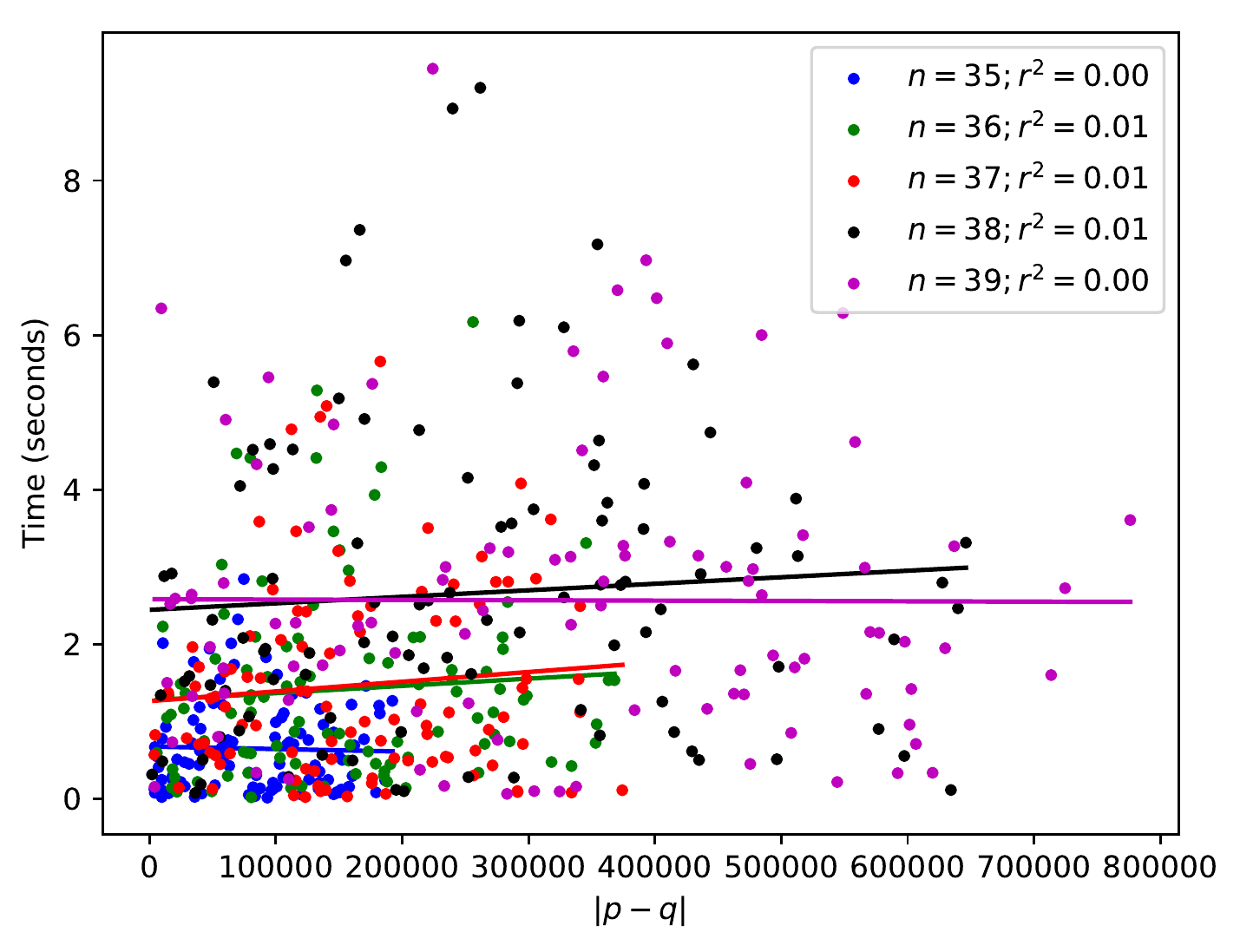}
  \caption{}\label{sfig:patlongdiff}                          
\end{figure}
\begin{figure}
  \includegraphics[width=\columnwidth]{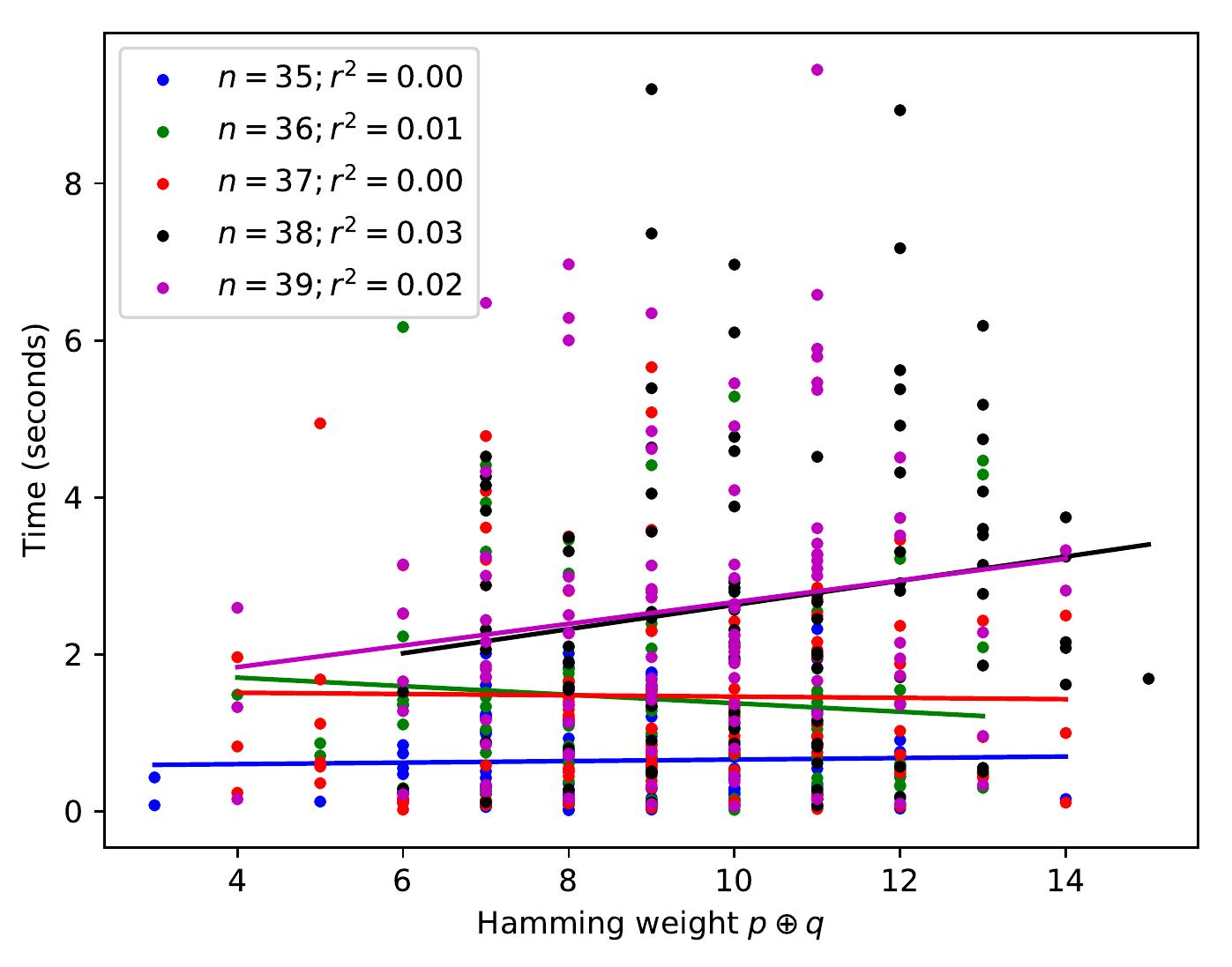}
  \caption{}\label{sfig:patlonghdpq}                          
\end{figure}
\begin{figure}
  \includegraphics[width=\columnwidth]{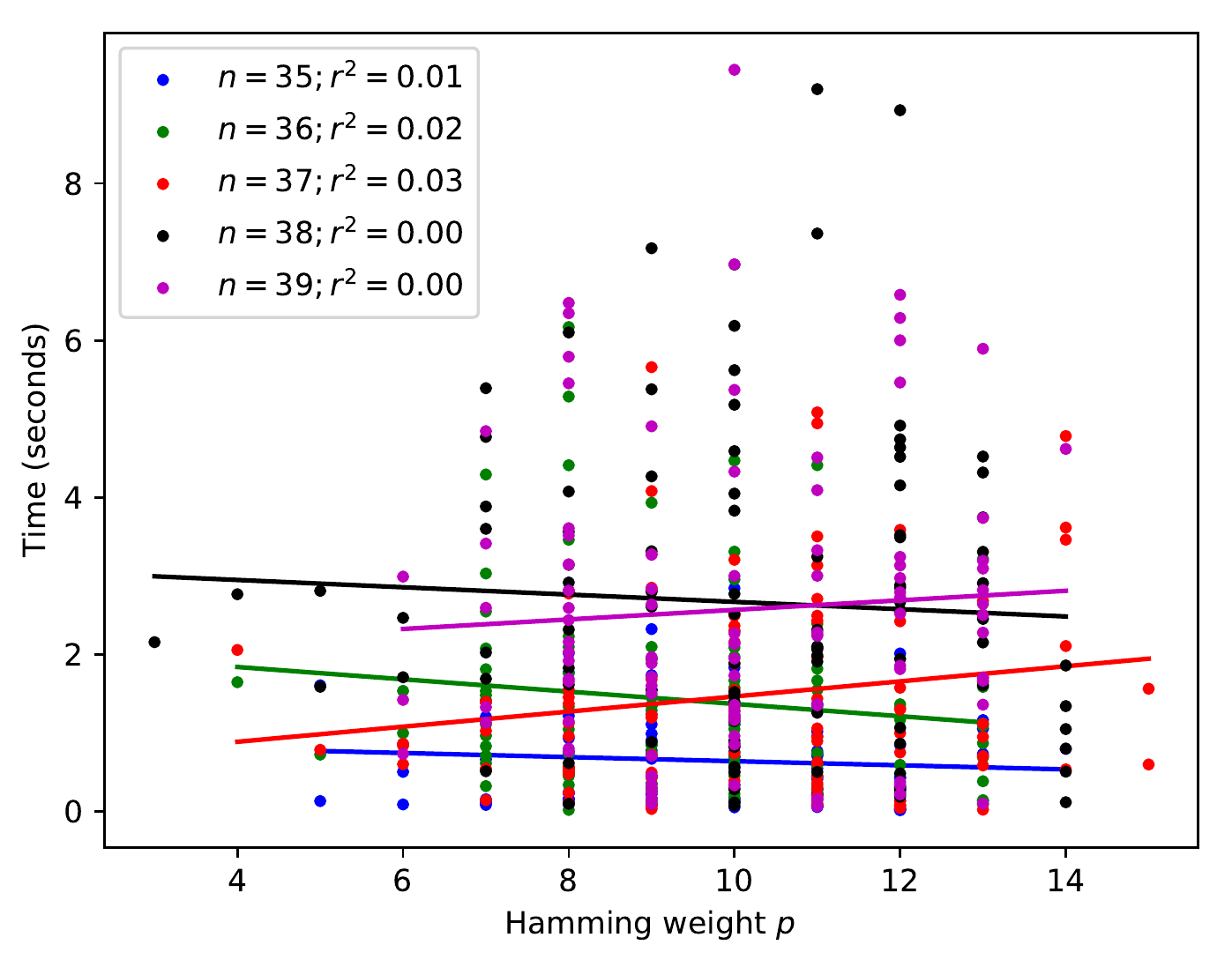}
  \caption{}\label{sfig:patlonghwp}                            
\end{figure}
\begin{figure}
  \includegraphics[width=\columnwidth]{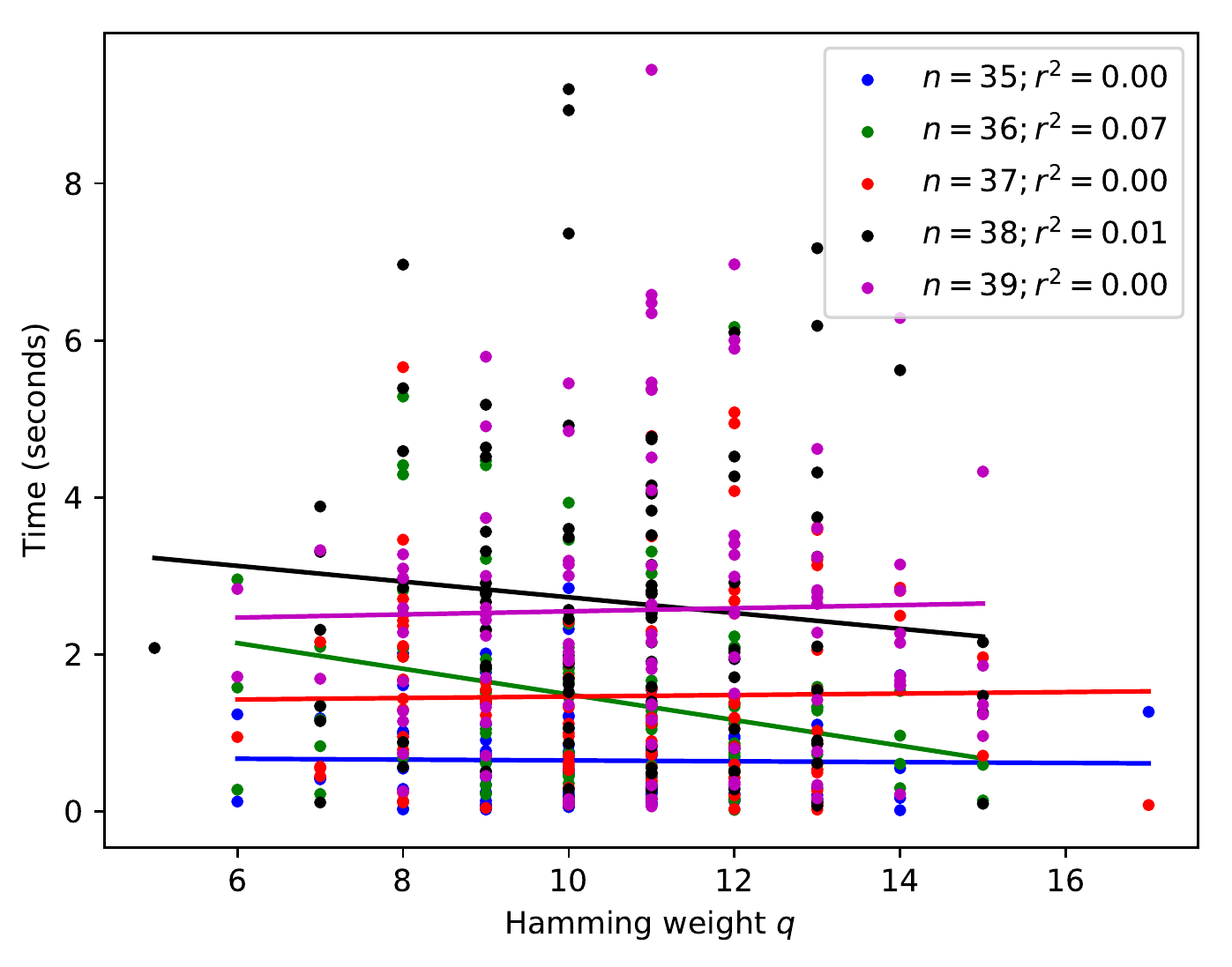}
  \caption{}\label{sfig:patlonghwq}
\end{figure}
\begin{figure}
  \includegraphics[width=\columnwidth]{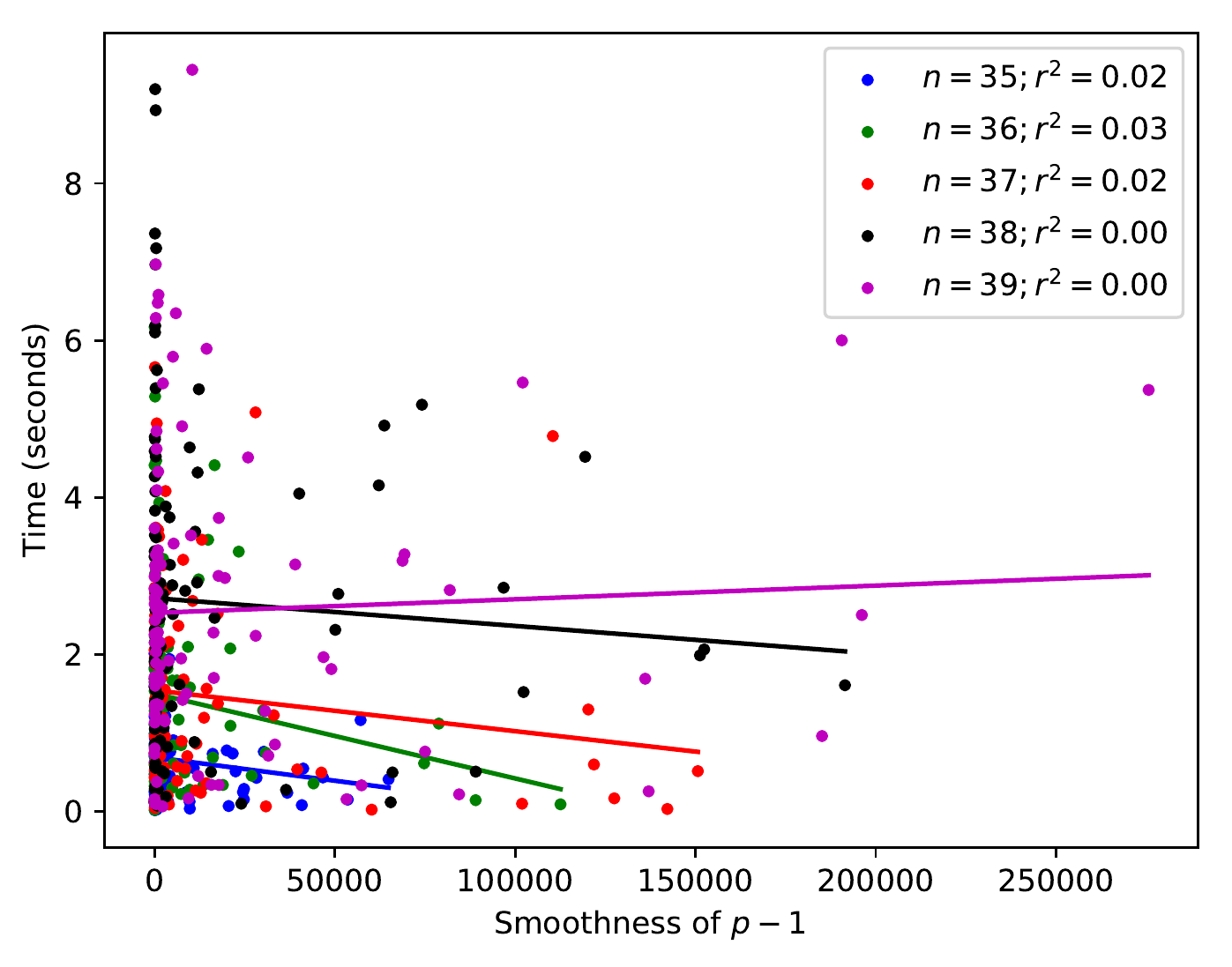}
  \caption{}\label{sfig:patlongsmp}                          
\end{figure}
\begin{figure}
  \includegraphics[width=\columnwidth]{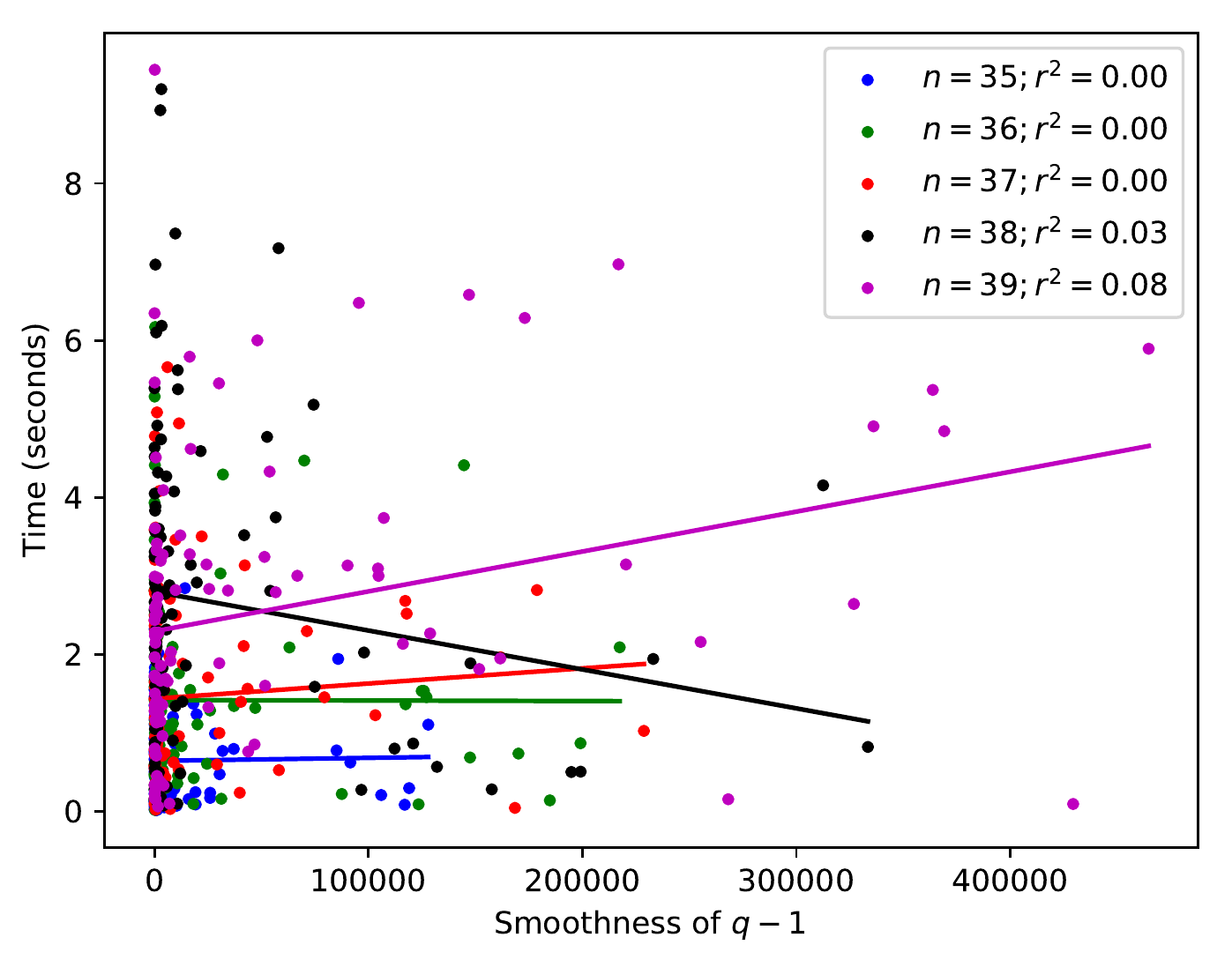}
  \caption{}\label{sfig:patlongsmq}                            
\end{figure}

\begin{figure}
  \includegraphics[width=\columnwidth]{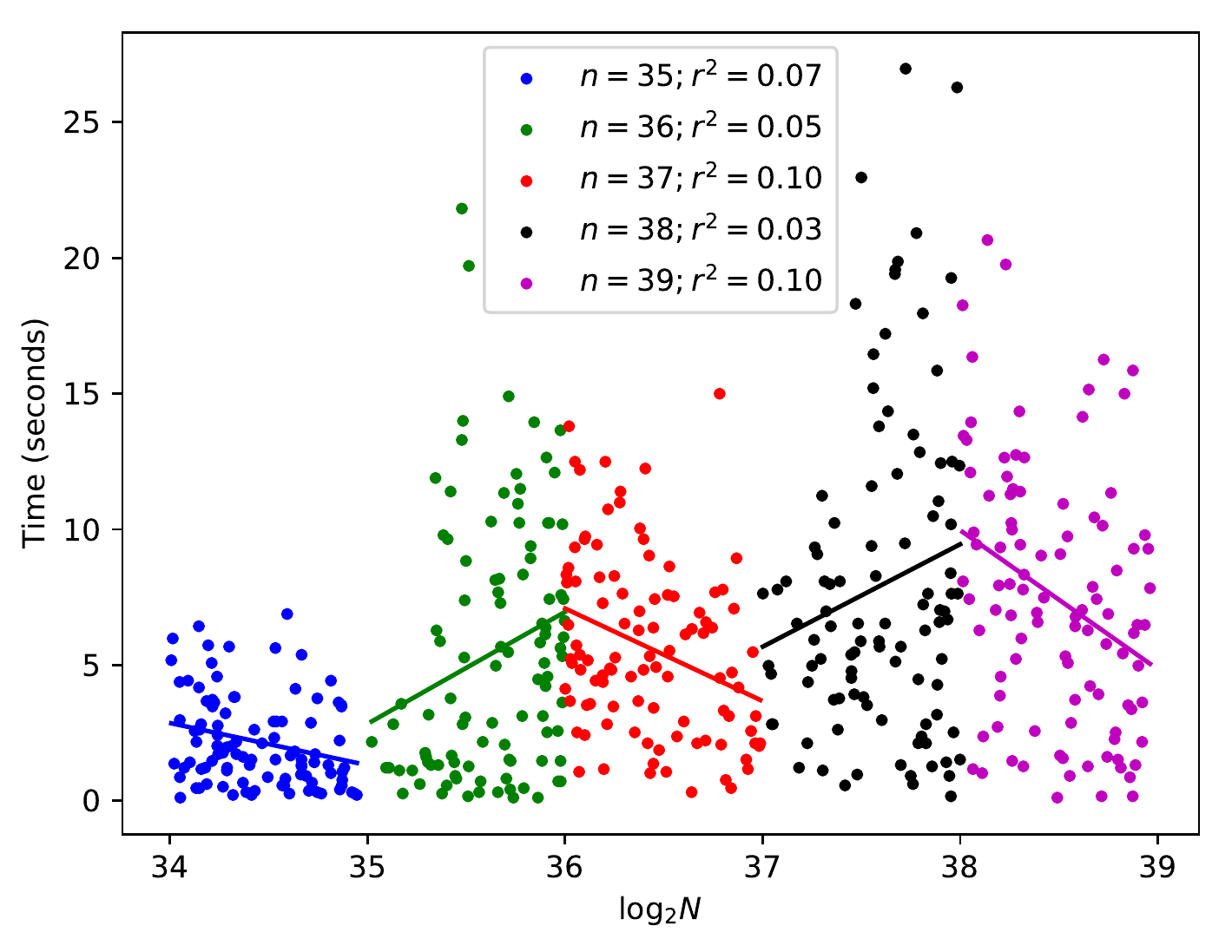}
  \caption{}\label{sfig:patkatsize}
\end{figure}
\begin{figure}
  \includegraphics[width=\columnwidth]{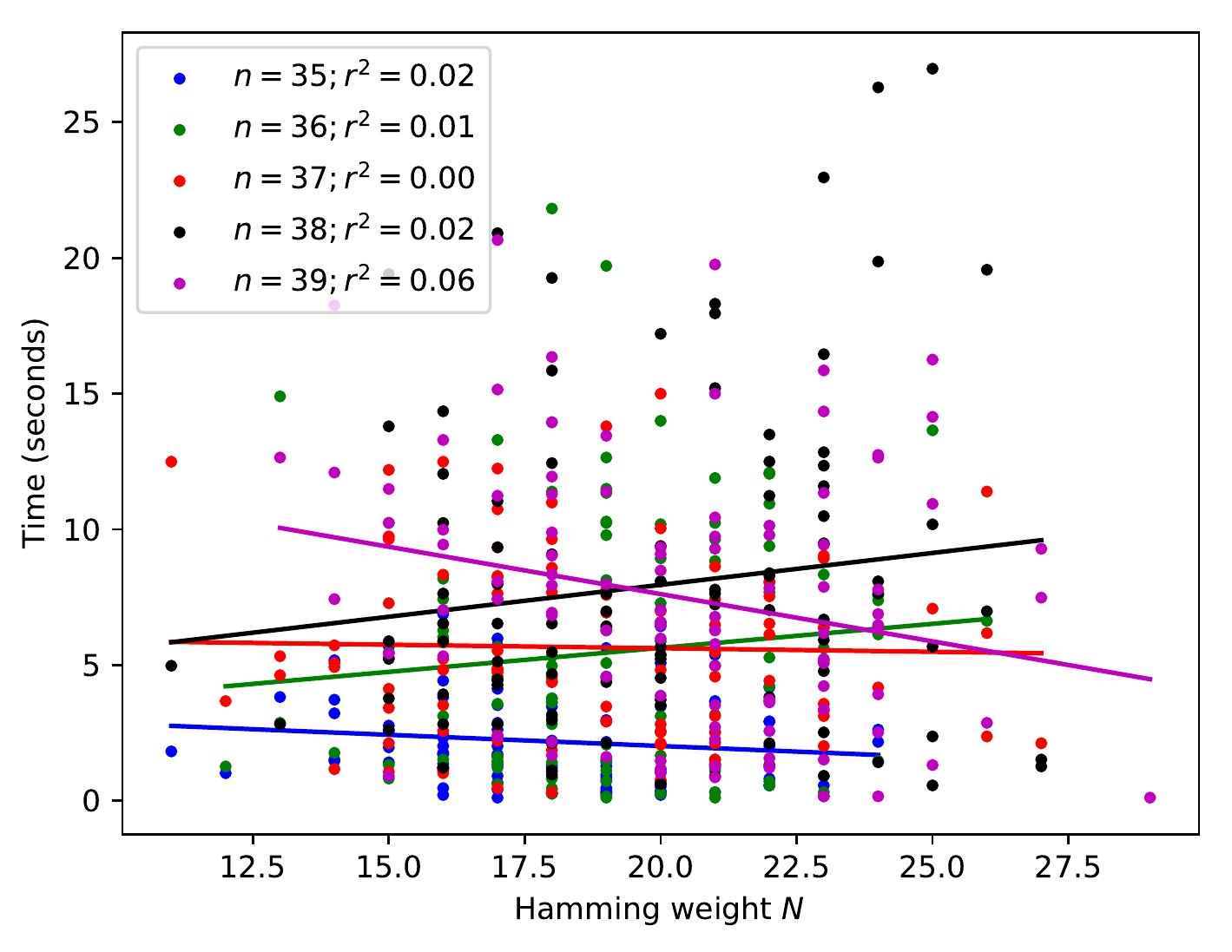}
  \caption{}\label{sfig:patkathwn}
\end{figure}
\begin{figure}
  \includegraphics[width=\columnwidth]{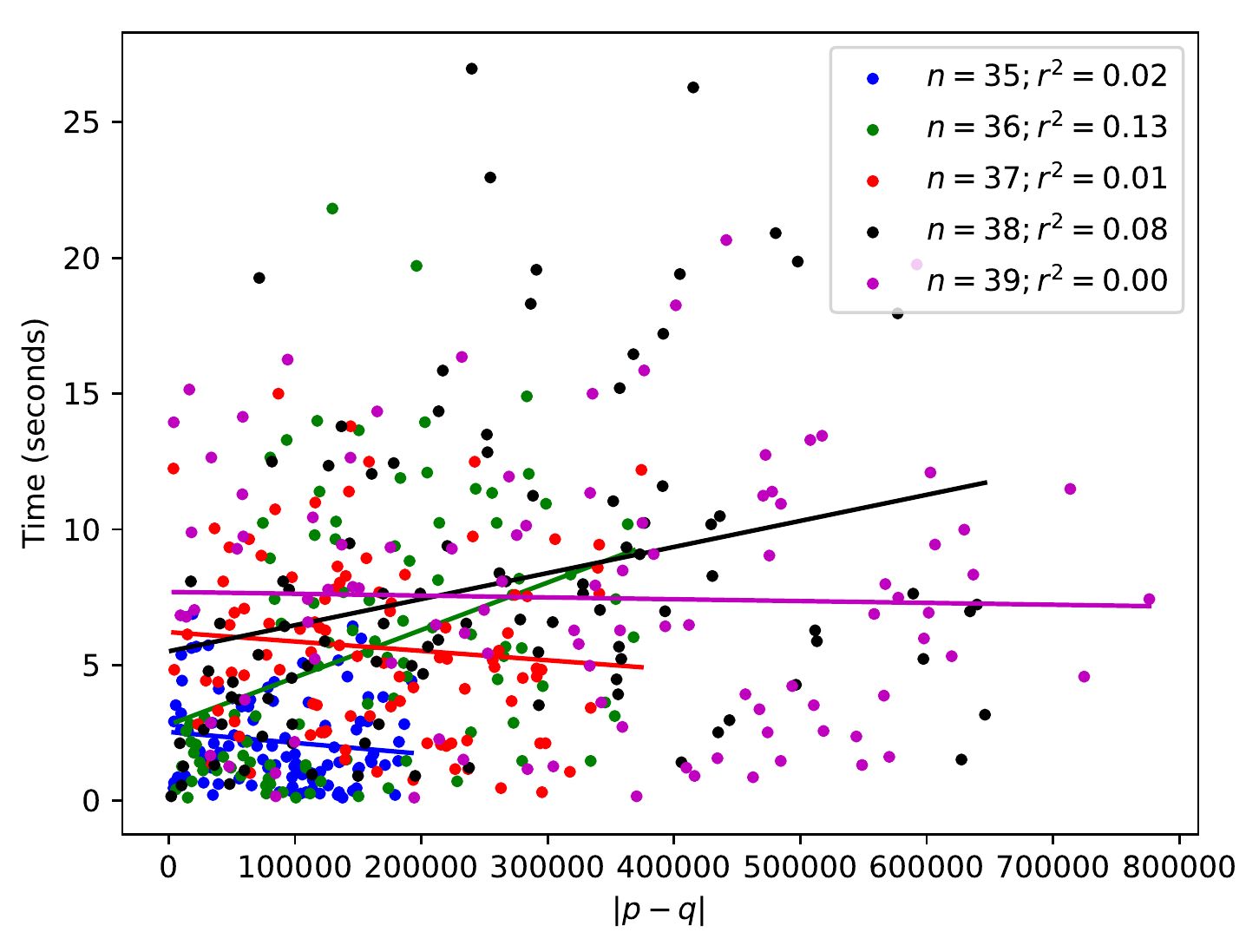}
  \caption{}\label{sfig:patkatdiff}
\end{figure}
\begin{figure}
  \includegraphics[width=\columnwidth]{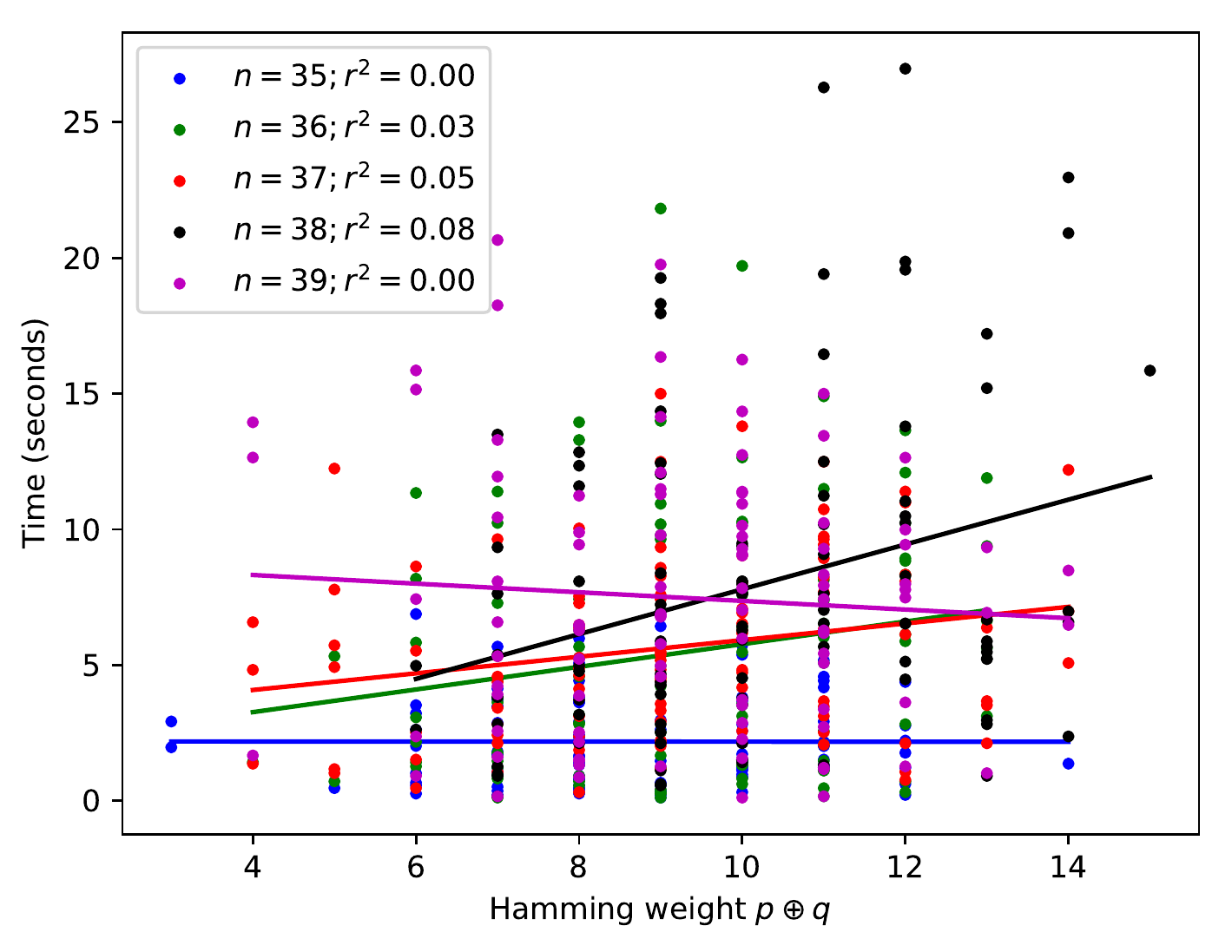}
  \caption{}\label{sfig:patkathdpq}
\end{figure}
\begin{figure}
  \includegraphics[width=\columnwidth]{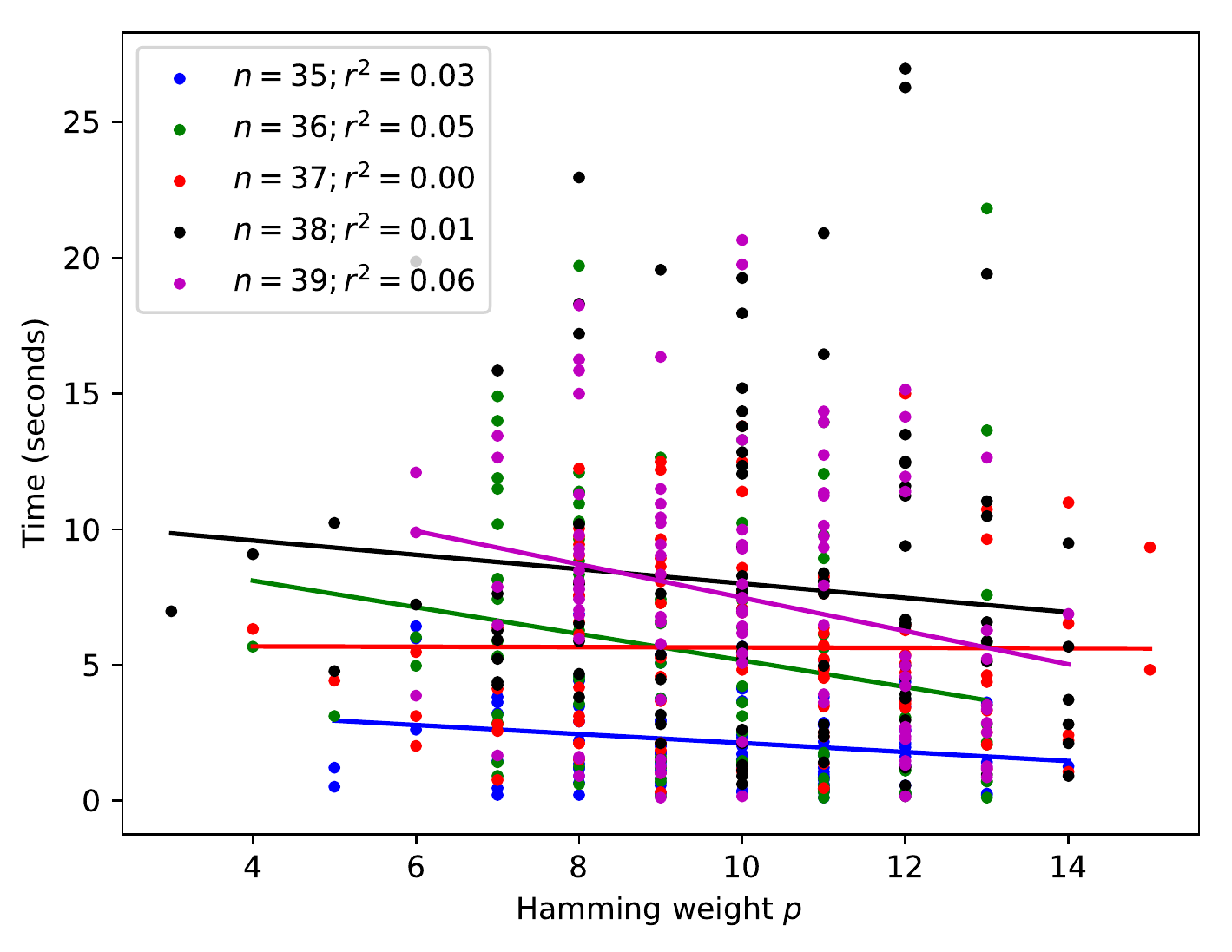}
  \caption{}\label{sfig:patkathwp}
\end{figure}
\begin{figure}
  \includegraphics[width=\columnwidth]{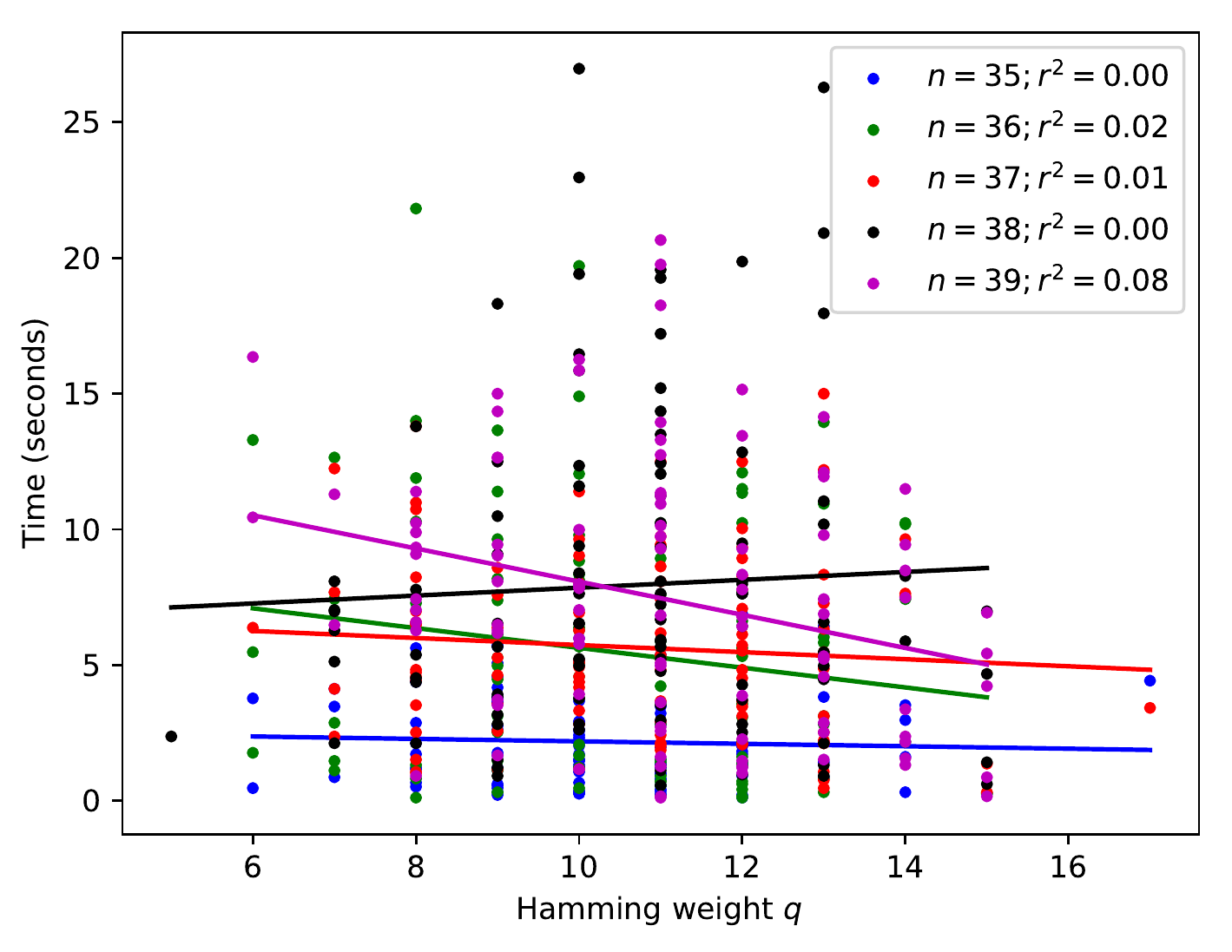}
  \caption{}\label{sfig:patkathwq}
\end{figure}
\begin{figure}
  \includegraphics[width=\columnwidth]{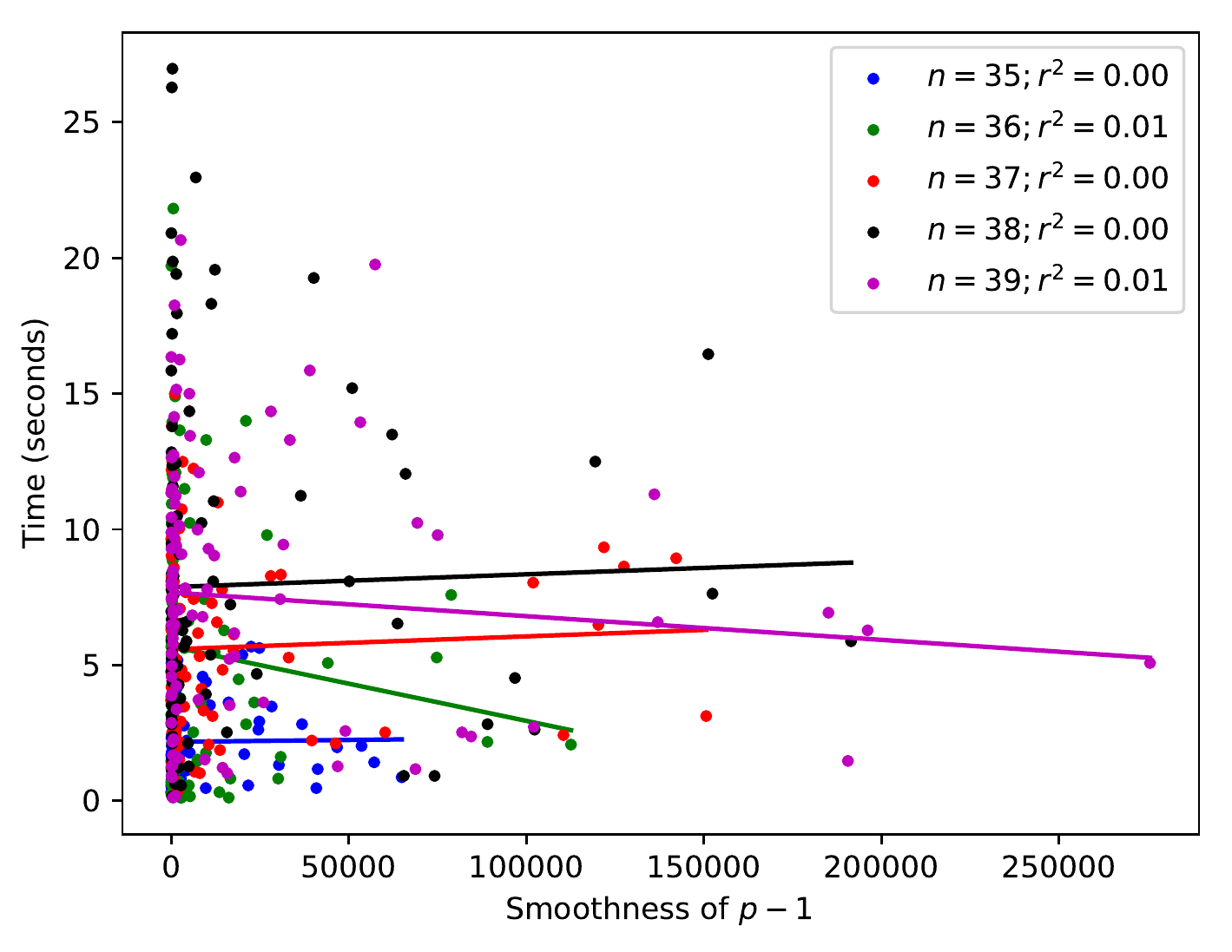}
  \caption{}\label{sfig:patkatsmp}
\end{figure}
\begin{figure}
  \includegraphics[width=\columnwidth]{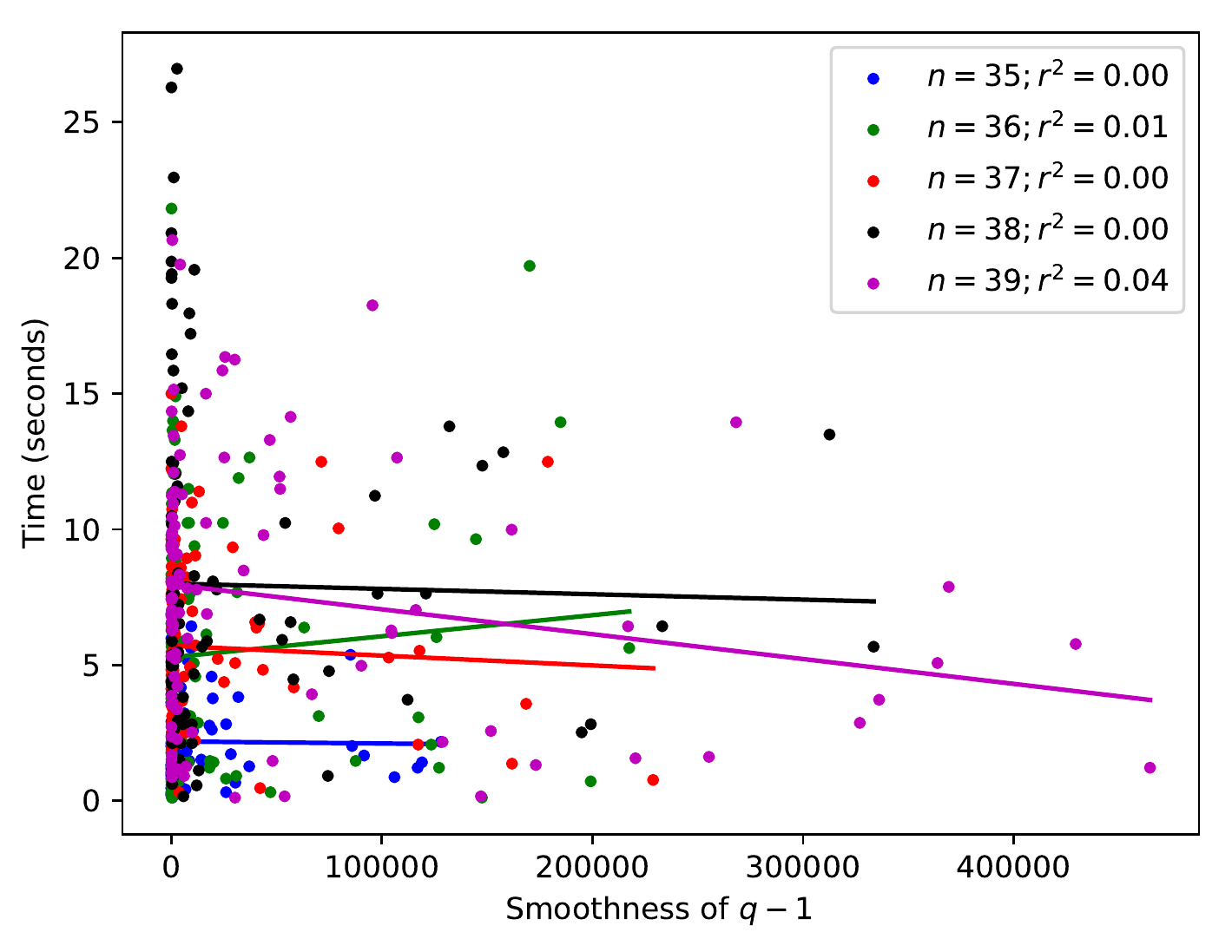}
  \caption{}\label{sfig:patkatsmq}
\end{figure}

\else 

\begin{figure}
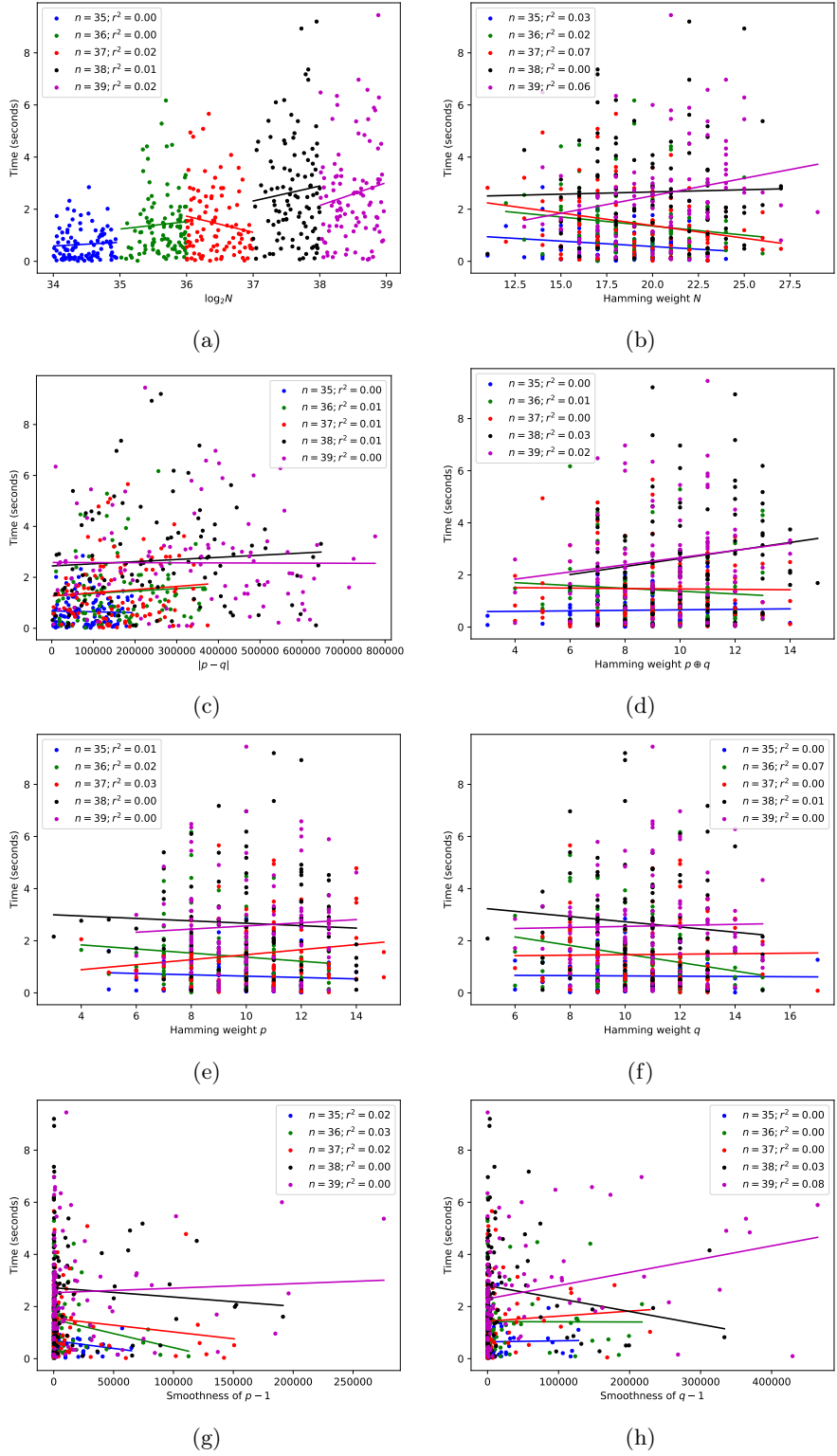

  \centering
  \begin{subfigure}[t]{0.37\textwidth}
    \includegraphics[width=\textwidth]{patterns/long_size.pdf}
    \caption{}\label{sfig:patlongsize}                           
  \end{subfigure}                                               
  ~                                                             
  \begin{subfigure}[t]{0.37\textwidth}                         
    \includegraphics[width=\textwidth]{patterns/long_hwN.pdf}
    \caption{}\label{sfig:patlonghwn}                          
  \end{subfigure}                                             
  
  \begin{subfigure}[t]{0.37\textwidth}                          
    \includegraphics[width=\textwidth]{patterns/long_diff.pdf}
    \caption{}\label{sfig:patlongdiff}                          
  \end{subfigure}                                              
  ~                                                            
  \begin{subfigure}[t]{0.37\textwidth}                        
    \includegraphics[width=\textwidth]{patterns/long_hdpq.pdf}
    \caption{}\label{sfig:patlonghdpq}                          
  \end{subfigure}                                            
                                                             
  \begin{subfigure}[t]{0.37\textwidth}                       
    \includegraphics[width=\textwidth]{patterns/long_hwp.pdf}
    \caption{}\label{sfig:patlonghwp}                            
  \end{subfigure}                                               
  ~
  \begin{subfigure}[t]{0.37\textwidth}                          
    \includegraphics[width=\textwidth]{patterns/long_hwq.pdf}
    \caption{}\label{sfig:patlonghwq}
  \end{subfigure}
                                                              
  \begin{subfigure}[t]{0.37\textwidth}                        
    \includegraphics[width=\textwidth]{patterns/long_smp.pdf}
    \caption{}\label{sfig:patlongsmp}                          
  \end{subfigure}                                            
  ~                                                           
  \begin{subfigure}[t]{0.37\textwidth}                       
    \includegraphics[width=\textwidth]{patterns/long_smq.pdf}
    \caption{}\label{sfig:patlongsmq}                            
  \end{subfigure}                                               
  \caption{Solver time versus various patterns (schoolbook encoding).}
  \label{fig:patlong}
\end{figure}

\begin{figure}
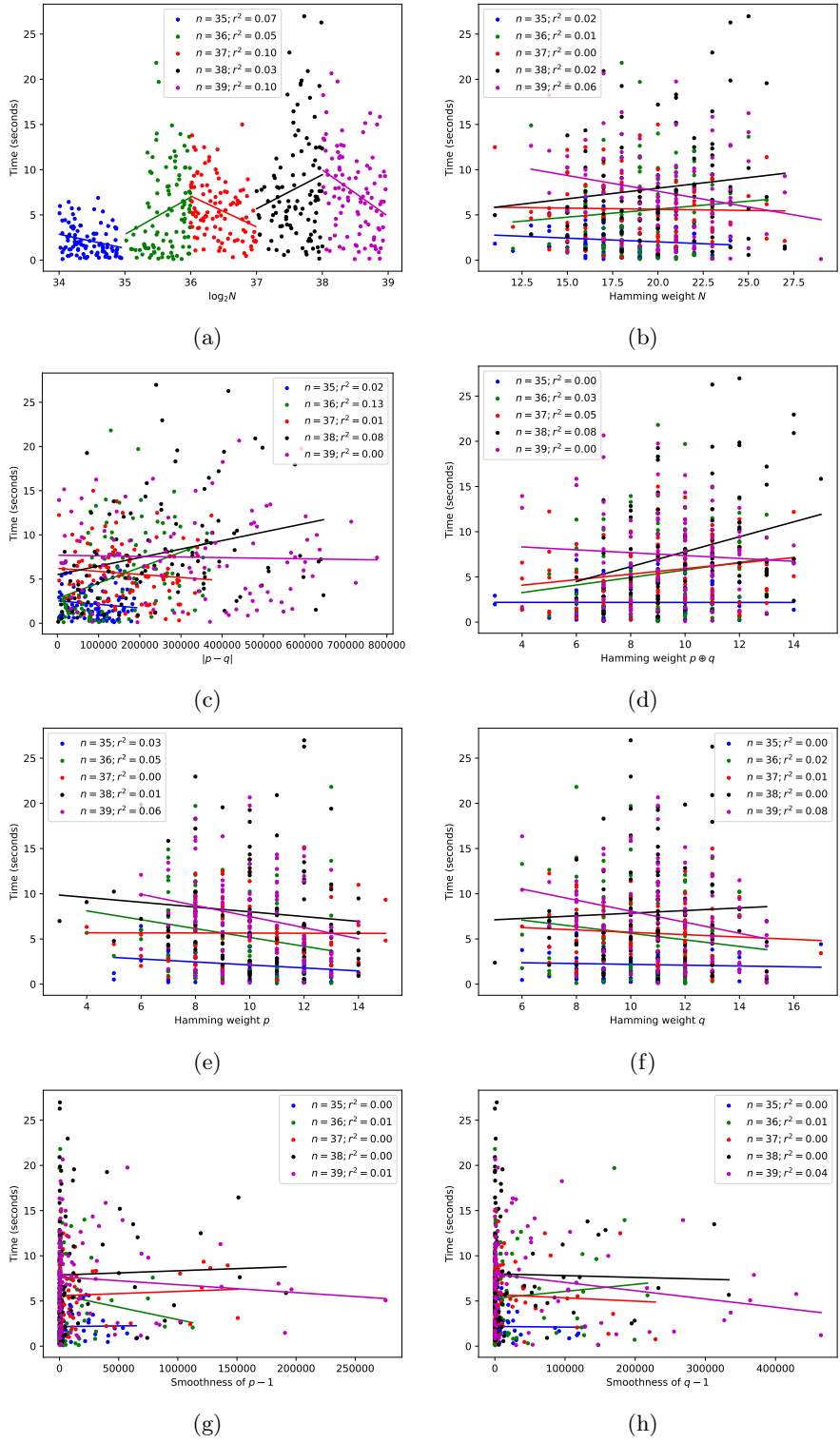

  \centering
  \begin{subfigure}[t]{0.37\textwidth}
    \includegraphics[width=\textwidth]{patterns/karatsuba_size.pdf}
    \caption{}\label{sfig:patkatsize}
  \end{subfigure}
  ~
  \begin{subfigure}[t]{0.37\textwidth}
    \includegraphics[width=\textwidth]{patterns/karatsuba_hwN.pdf}
    \caption{}\label{sfig:patkathwn}
  \end{subfigure}
  
  \begin{subfigure}[t]{0.37\textwidth}
    \includegraphics[width=\textwidth]{patterns/karatsuba_diff.pdf}
    \caption{}\label{sfig:patkatdiff}
  \end{subfigure}
  ~
  \begin{subfigure}[t]{0.37\textwidth}
    \includegraphics[width=\textwidth]{patterns/karatsuba_hdpq.pdf}
    \caption{}\label{sfig:patkathdpq}
  \end{subfigure}
  
  \begin{subfigure}[t]{0.37\textwidth}
    \includegraphics[width=\textwidth]{patterns/karatsuba_hwp.pdf}
    \caption{}\label{sfig:patkathwp}
  \end{subfigure}
  ~
  \begin{subfigure}[t]{0.37\textwidth}
    \includegraphics[width=\textwidth]{patterns/karatsuba_hwq.pdf}
    \caption{}\label{sfig:patkathwq}
  \end{subfigure}
  
  \begin{subfigure}[t]{0.37\textwidth}
    \includegraphics[width=\textwidth]{patterns/karatsuba_smp.pdf}
    \caption{}\label{sfig:patkatsmp}
  \end{subfigure}
  ~
  \begin{subfigure}[t]{0.37\textwidth}
    \includegraphics[width=\textwidth]{patterns/karatsuba_smq.pdf}
    \caption{}\label{sfig:patkatsmq}
  \end{subfigure}
  \caption{Solver time versus various patterns (Karatsuba encoding).}
  \label{fig:patkat}
\end{figure}

\fi 

\ifacm
We examined
\else 
\autoref{fig:patlong} and \autoref{fig:patkat} examine
\fi 
the relation between various metrics on $p$, $q$ and the solver time for
\ifacm
  long multiplication (Figures~\ref{sfig:patlongsize}--\ref{sfig:patlongsmq})
  and Karatsuba encoding (Figures~\ref{sfig:patkatsize}--\ref{sfig:patkatsmq}).
\else 
long multiplication encoding and Karatsuba encoding (respectively).
\fi 
See also~\citerepo{} for enlarged images.
We examined bitwise patterns as these are most likely exploited
by the SAT solver and we examined smoothness as this can determine
the hardness of factoring for some number-theoretical methods.

Note that only the first two metrics ($\log_2 N$ and Hamming
weight$(N)$) could potentially be used to predict how fast the solver
will find a solution.  The remaining metrics require knowledge of the
value of $p$ and $q$, but these metrics could be important for anyone
generating primes in the RSA cryptosystem.

However, the lack of any correlation indicates that none of the
investigated patterns have a significant impact on the solver
time.  In other words, SAT solvers do not influence the method by
which a user of the RSA cryptosystem should generate primes.

\end{document}